\providecommand{\columnwiselinenumbersfalse}{}
\providecommand{\runningpagewiselinenumbers}{}
\providecommand{\linenumbers}{}

\documentclass[desactivate]{aa} 

\usepackage{graphicx}
\usepackage{txfonts}
\usepackage{hyperref}
\usepackage{rotating}
\usepackage{xcolor}

\graphicspath{{./images/}{../images/}}

\newcommand{\urad}{w_\text{rad}}

\begin{document} 

\title{High-resolution simulations of non-thermal emission from LS~5039}

\author{
   R. Kissmann\inst{1}
   \and
   D. Huber\inst{1}
   \and
   P. Gschwandtner\inst{2}
}

\institute{
Universit\"at Innsbruck,
Institut f\"ur Astro- und Teilchenphysik,
Technikerstr. 25,
6020 Innsbruck, Austria~\\
\email{ralf.kissmann@uibk.ac.at}
\and
Universit\"at Innsbruck,
Institut f\"ur Informatik,
Technikerstr. 21 a,
6020 Innsbruck, Austria
}

\abstract{
In a previous study, we investigated the relativistic wind dynamics in the LS 5039 system.
In this work, we analyse energetic-particle transport within this modelling context, where we simulate the high-energy particle distribution and ensuing emission of non-thermal radiation.
}{
From these high-resolution simulations covering three full orbits, we compute the non-thermal emission from this system and compare it to corresponding observations.
}{
We modelled the LS 5039 system assuming a wind-driven scenario.
Our numerical model uses a joint simulation of the dynamical wind interaction together with the transport of energetic leptons from the shocked pulsar wind.
We computed the non-thermal emission from this system in a post-processing step from the resulting distribution of energetic leptons.
In this computation, we took into account the synchrotron and inverse Compton emission, relativistic beaming, and $\gamma\gamma$-absorption in the stellar radiation field.
}{
We investigated the dynamical variation of the energetic particle spectra on both orbital and on short timescales.
From our simulation of three full orbits,  we were also able to investigate the orbit-to-orbit variability.
Our model successfully reproduces many of the spectral features of LS 5039. 
We also find a better correspondence between our predicted orbital light curves and the corresponding observations in soft x-rays, low-energy, and high-energy gamma rays than in our previous modelling efforts.
}{
We find that our high-resolution and large-scale simulations can successfully capture the relevant parts of the wind-collision region that are related to particle acceleration and emission of non-thermal radiation.
The quality of the fit strengthens the wind-driven assumption underlying our model.
Desirable extensions for the future include a dynamical magnetic-field model for the synchrotron regime, a revision of our injection parameters,  and a consideration of an additional hadronic component that could explain recent observations in the 100~TeV regime.
}


\maketitle

\section{Introduction} \label{sec_intro}
Gamma-ray binaries are binary systems that emit the majority of their radiation in the gamma-ray regime.
They are composed of an early-type star and a compact object, either a neutron star or a black hole.
There are two prominent scenarios to explain their emission. On the one hand, we have a microquasar scenario, where a relativistic jet related to the compact object is assumed to be the site of non-thermal particle acceleration \citep[see e.g.][]{Bosch-RamonKhangulyan2009IJMPD18_347}. On the other hand, in a wind-driven scenario, the compact object is assumed to be a pulsar and the particles are accelerated (e.g. at shocks in the wind-collision region, WCR) of the stellar and the pulsar wind \citep[see e.g.][]{MaraschiEtAl1981MNRAS194_1, Dubus2006AnA456_801}. Correspondingly, a distinction between these scenarios can be helped by a clear identification of the compact object.
Specifically, presence of a black hole points towards a microquasar, while a pulsar indicates a wind-driven scenario.

In this study, we investigate the LS 5039 system, a gamma-ray binary that has been extensively studied in a broad wavelength range.
Its obvious orbital variability in different wavelength regimes \citep[][]{YonedaEtAl2021ApJ917_90, CollmarZhang2014AnA565_38, Aharonian2005Sci309_746} motivates a wind-driven scenario.
However, the true nature of this compact object has not been clearly identified thus far.
So far, there are only hints to possible pulsations in X-rays \citep[see][]{YonedaEtAl2020PhRvL125_1103, Volkov2021ApJ915_61}

In the wind-driven scenario, most models assume the non-thermal emission from LS-5039 to have a purely leptonic origin.
Correspondingly, the dominant emission processes are synchrotron and inverse-Compton (IC) emission \citep[][]{Zabalza2013AnA551_17, DubusEtAl2015AnA581_27, MolinaBosch-Ramon2020AnA641_84}.

In particular, for non-thermal particles transported in a relativistic outflow, there is a multitude of effects that lead to an orbital variation.
Whereas synchrotron emission would show orbital variation from variation in the energetic particle density and the magnetic field (energy density and direction), IC scattering is inherently anisotropic, depending on the relative position of the source of the scattered radiation and the observer.
With the relevant photons stemming from the stellar companion, IC scattering is most efficient at superior conjunction.
However,  for large inclination angles in particular, the very-high energy (VHE) gamma-ray flux is attenuated due to pair-production with the stellar wind photons,  thus acting against the increase in VHE flux due to more efficient IC scattering.
Finally, the emission is further modulated by relativistic boosting in the shocked, relativistic pulsar wind \citep[][]{Bogovalov2009MNRAS387_63, DubusEtAl2010AnA516_18}, which depends on the alignment of the relativistic flow with the direction towards the observer.

These combined effects are thought to explain the energy-dependent orbital variability of LS-5039.
In particular, observations show an anti correlation of the high-energy (HE) gamma-ray emission with the x-ray and VHE gamma-ray emission.
Due to the complex interplay of the different orbital effects, a one-zone model cannot explain the light curve at different energies.
Therefore, models aimed at explaining the non-thermal emission from this system use extended emission regions or multi-zone models \citep[see e.g.][]{Bosch-RamonBarkov2011AnA535_20, Zabalza2013AnA551_17}.
In \citet{Zabalza2013AnA551_17}, the authors discuss the Coriolis shock as a possible site for the emission of VHE photons because reduced energy losses as compared to the shock at the apex of the WCR should allow particles to reach higher energies.

These studies also show the necessity to take the full complexity of the wind interaction into account, when aiming at a detailed comparison to observations.
The dynamical wind interaction has been studied in dedicated numerical simulations using a description via relativistic hydrodynamics \citep[RHD, see][]{LambertsEtAl2013AnA560_79,Bosch-RamonEtAl2015AnA577_89,  KissmannEtAl2023AnA677_A5} with even first relativistic magneto-hydrodynamics models recently appearing \citep[][]{BarkovEtAl2024PASA41_48}.
Such more detailed wind models allowed us to obtain a more complex description for the emission model by computing the non-thermal emission in a post-processing step for a given orbital state \citep[][]{DubusEtAl2015AnA581_27, MolinaBosch-Ramon2020AnA641_84}.

A first model coupling wind dynamics and particle transport was presented by \citet{HuberEtAl2021AnA646_A91} which allowed for the particle distribution to react to dynamical effects of the plasma.
This simulation led to a good agreement between model and observations regarding large parts of the spectral energy distribution (SED) and the orbital variation.
However, the corresponding study of LS-5039 \citep{HuberEtAl2021AnA649_71} also suffered from the deficiency that the numerical domain was too small to capture the Coriolis shock at all orbital phases.
Therefore, we devised a new simulation, which features both a significantly larger domain and also increased spatial resolution.

The dynamics of the interacting pulsar and stellar winds of this new model is discussed in \citet{KissmannEtAl2023AnA677_A5}.
In this case, the Coriolis shock and the unshocked pulsar wind were captured by the simulation at all times.
Additionally, the higher spatial resolution led to stronger turbulence in the simulation, which resulted in considerable dynamics over a full orbit and even distinct differences from orbit to orbit.

Here, we use the same dynamical, high-resolution simulations in conjunction with a numerical model for the transport of energetic particles as discussed in \citet{HuberEtAl2021AnA646_A91}.
Correspondingly, we were able to model the dynamical evolution of the non-thermal particles within the simulation domain.
Additionally, we computed the non-thermal emission of these energetic particles as a function of energy, orbital phase, and observer direction.

The paper is structured as follows.
After this introduction, we present the specifics of our numerical model together with the related mathematical description in Sect. \ref{SecSetup}.
The results of the simulations for the energetic particles and the non-thermal emission are discussed in Sect. \ref{SecResults}.
Finally, we finish with a summarising discussion in Sect. \ref{SecConclusion}.

\section{Physical and numerical setup}
\label{SecSetup}
LS-5039 is a binary system consisting of a massive O-type star and a compact object, which we assume to be a pulsar in this study.
The setup of the orbit and the properties of the stellar components have been adapted from \citet{CasaresEtAl2005MNRAS364_899, DubusEtAl2015AnA581_27}.
We modelled the dynamical interaction between the stellar and the relativistic pulsar wind using an RHD description as detailed in \citet{KissmannEtAl2023AnA677_A5}.
These simulations used a homogeneous spatial resolution of $\sim0.42R_\odot$ on a grid with $N_x \times N_y \times N_z  = 2048 \times 1536 \times 1024$ cubic cells, corotating with the average angular velocity of the system.
We solved the RHD equations with a relativistic extension of the \textsc{Cronos} code \citep[][]{KissmannEtAl2018ApJS236_53, HuberKissmann2021AnA653_164}.
These simulations were run for three full orbits, where the first orbit was used to allow the system to evolve into the fully turbulent state.

In the present study, we focus on the transport of energetic particles within these interacting winds and on the related gamma-ray emission.
For this, we additionally solved the transport equation of the energetic particles alongside the fluid simulation, where the velocity of the gas from the fluid simulation was used as the advection velocity of the particles.
Finally, the ensuing emission from this system was computed in a post-processing step from the particle distribution together with the dynamically simulated fluid properties, with the latter needed for relativistic boosting effects.
In the following, we give some details on modelling the particle transport and the emission of non-thermal radiation.

\subsection{Particle transport}
\label{SecInjection}
The full description of our particle-transport scheme can be found in \citet{HuberEtAl2021AnA646_A91}.
Here, we only give an overview of the implementation and discuss those details relevant for the present context.
In particular, we solve the following transport equation for  the
number density of electrons and positrons, $n_l'$, 
\begin{equation}
  {\nabla_\mu \left( u^\mu n_l' \right)}
  + \frac{\partial}{\partial \gamma'} 
  \left(
      \left\langle \dot{\gamma}' \right\rangle
    n_l'
    \right)
  = 0, \label{eq: transport final}
\end{equation}
where $N = n_l' \text{d}^3x' \text{d}\gamma'$ is the number of leptons in the spatial volume, $\text{d}^3x'$, and in the interval, $[\gamma',\gamma'+ \text{d}\gamma']$, with $\gamma'$ the Lorentz factor of the particles.
Additionally, $\left\langle \dot{\gamma}' \right\rangle$ represents the total rate of energy change in the fluid frame. 
For this, we take three energy-loss processes into account: 
adiabatic energy changes and synchrotron and IC energy losses.
In contrast to transport in a non-relativistic fluid, adiabatic energy changes in a relativistic fluid depend on the four-dimensional divergence of the relativistic four velocity (i.e. the divergence also depends on the rate of change of the local Lorentz factor of the fluid).
To compute this rate of change, we stored the Lorentz factor $\gamma(t^{n-1})$ from the previous time step and then used the approximation,
\begin{equation}
        \frac{\partial \gamma}{\partial t} \simeq \frac{\gamma(t^{n}) - \gamma(t^{n-1})}{\Delta t}.
\end{equation}
Since in our simulation velocities are given in units of the speed of light, we also have $\gamma = \sqrt{1+\vec{u}^2}$ with $\vec{u}$ the spatial part of the fluid's relativistic four velocity.

The IC losses are computed using the Klein-Nishina factor from \citet{Moderski2005MNRAS363_954} with the radiative energy density from the stellar radiation field of the massive stellar companion given by
\begin{equation}
  \urad = \frac{L_\mathrm{star}}{4 \pi r_\mathrm{star}^2 c},
\end{equation}
where we used $L = 1.8\times 10^5 L_{\sun}$ \citep[see][]{CasaresEtAl2005MNRAS364_899}.

The synchrotron energy-loss rate depends on the local magnetic field.
Since our current simulations are based on an RHD model, the magnetic field does not follow directly from the fluid simulation, but we have to rely on a model for the magnetic field instead.
As also discussed in \citet{HuberEtAl2021AnA646_A91}, we assume the magnetic-field energy density to be a fraction, $\zeta_b=0.5$, of the internal energy.

\subsection{Particle injection}
We injected energetic leptons at strongly compressive regions in the interacting stellar winds, as described in \citet{HuberEtAl2021AnA646_A91}.
Also motivated by \citet{DubusEtAl2015AnA581_27}, we injected a combination of a Maxwellian distribution of particles heated at the shocks and a power law representing the non-thermal particles.
The extent and normalisation of the power law were fixed by prescribing the fractions of the number density  and thermal energy density of the pair plasma to be transformed into non-thermal particles, $\zeta_n^{\text{PL}}$ and $\zeta_e^{\text{PL}}$, respectively.
Additionally, we used a spectral index of $s=1.5$ for the injected power law, namely, using the same values as in our previous study.
Here, we ensured that only the density related to the pulsar wind was taken into account in producing the 
non-thermal particle population.
The rest of the number density and the thermal energy density is assigned to the Maxwellian, representing the thermal part of the pair plasma.

In our simulation, the highly relativistic pulsar wind is modelled with a Lorentz factor only slightly larger than 7, corresponding to an outflow with 0.99\,c, whereas the Lorentz factor for LS 5039 could be as high as 10$^4$. 
Since we used a spin-down luminosity of $L_\text{sd} = 7.55\times 10^{28}$ W, we correspondingly expected to significantly overestimate the number density of the pulsar wind.
Therefore, we only inject a fraction, $\zeta_\rho$, of the pair-plasma number density into the Maxwellian and the power law.

This fraction of the pair-plasma number density is distributed between the Maxwellian and the power law, with the fraction $\zeta_n^\text{PL}$ attributed to the power law.
Additionally, the thermal energy density of the pair plasma is distributed between these two particle populations, with $\zeta_e^\text{PL}$ indicating the fraction assigned to the power law. 
On the one hand, only a small fraction of the particles of the pair-plasma will be converted to non-thermal particles; namely, we have $\zeta_n^\text{PL} \ll 1$ and for the fraction of particles attributed to the Maxwellian, we have $\zeta_n^\text{MW} = 1 - \zeta_n^\text{PL} \simeq 1$.
On the other hand, the non-thermal particles contain a major fraction of the available energy density. 
Since, $\zeta_\rho$ scales the density to be converted to any of the two populations, these scaling parameters determine the normalisation and extent in energy of both populations in a rather complex fashion \citep[for further details, see][]{HuberEtAl2021AnA646_A91}.

Because the exact Lorentz factor of the pulsar wind is unknown, the $\zeta_\rho$ parameter (and, correspondingly, the other scaling parameters as well) has to be adapted according to observations of the non-thermal emission.
However, tuning these parameters during the present large-scale simulations was not possible due to the high computational cost.
Therefore, we used the same set of parameters as in \citet{HuberEtAl2021AnA649_71}, namely: $\zeta_n^{\text{PL}} = 4\cdot 10^{-3}$, $\zeta_e^{\text{PL}} = 0.45$, and $\zeta_\rho = 5.5\cdot 10^{-4}$.
These parameters also determine the extent of the power law of the non-thermal leptons.
Their maximum Lorentz factor follows from the balance of acceleration rate and synchrotron losses, assumed to dominate at the highest energies.
The minimum Lorentz factor follows by equating the integrated number density and energy density of the non-thermal leptons to the number density and energy density assigned to the power law \citep[for details, see also][]{HuberEtAl2021AnA646_A91}.

In the present analysis, we found that this specific set of parameters led to a higher non-thermal emission than in \citet{HuberEtAl2021AnA649_71}.
Given the complex dependence of the injected particle distribution on the scaling parameters, changing these parameters in the post-processing stage is not easily possible any more.
Here, we also have to bear in mind that the distinction between Maxwellian and non-thermal power law was only done for injection, while our working quantity $n_l'$ represents their joined distribution.
Nonetheless,  we still used a simple scaling of the particle spectra to improve the fit of the non-thermal emission spectrum,  as described below.
We also discuss the expected impact of a modified injection on other aspects of the spectrum, especially on the properties of the Maxwellian, along with our investigation of the details of the ensuing non-thermal emission spectrum.

Due to the short timescale for the population of energetic particles to reach a quasi-steady state and due to the huge demand of computational resources for a simulation including particles, we did not run the simulation over three full orbits including the energetic particles.
As also discussed in \citet{HuberEtAl2021AnA649_71} we restarted the RHD simulations at several orbital phases, adding the non-thermal particles and letting the simulation run for a physical timescale of approximately 80 minutes (such a single particle run required approximately 580000 core hours).
Then, we stored six output steps evenly spaced over a little more than the second half of this timescale to be used for the further analysis.

\subsection{Non-thermal emission}
The non-thermal emission from the energetic particles is simulated in a post-processing step from the energetic particle distribution within their dynamical environment.
The details of the corresponding computations are given in \citet{HuberEtAl2021AnA646_A91}.

Just as for the implementation of synchrotron losses, we also require a model for the magnetic field when computing synchrotron emission, where we use the same magnetic-field model as detailed above.
With this magnetic-field model, we only describe the amplitude of the field, but not its individual vector components.
This can also be viewed as approximating the magnetic field to be disordered.
In this case, we used the approximation $B_\perp' \simeq \frac{\pi}{4} B'$, namely, we assumed the perpendicular component to be equivalent to the pitch-angle average of the field.

When computing IC emission, we followed the prescription by \citet{Moderski2005MNRAS363_954}, where we used the scattering rate by \citet{AharonianAtoyan1981ApnSS79_321}.
For this, we used the photon field of the O-type star as the target-photon field, which we approximate as being monochromatic in the computation of IC emission.

Since the non-thermal particles are accelerated from the pulsar-wind material, they often propagate through regions with a relativistic plasma flow.
Therefore both emission processes are subject to relativistic beaming.
To take this into account, we computed the emission in the co-moving frame of the fluid and then transformed the emission into the observer frame by applying the corresponding Doppler boost \citep[see][for details of the computation]{HuberEtAl2021AnA646_A91}.
This can produce a significant orbital variability on top of the inherent one of both emission processes.

\section{Results}
\label{SecResults}
In this study, we modelled the dynamics and emission of energetic leptons in conjunction with the dynamical fluid of the interacting winds of the early-type star and the pulsar.
While the energetic particles are affected by the dynamical winds especially due to the dynamically changing fluid velocity, we did not consider a backreaction of the energetic particles on the fluid.
Therefore, we separately study the fluid dynamics of the given simulations in \citet{KissmannEtAl2023AnA677_A5}, where we discuss effects of short-term and long-term dynamics.
Here, we only investigate the distribution of energetic particles and the related non-thermal emission, without any further discussion of the fluid dynamics.

\subsection{Energetic particles}

\begin{figure*} 
  \setlength{\unitlength}{0.0175268cm}
  \begin{picture}(1027,1150)(0,0)

    \put(2,855){\rotatebox{90}{$y/\text{AU}$}}
    \put(2,1020){\rotatebox{90}{$y/\text{AU}$}}
    \put(800,1139){$n_l'(E = 5467.7\ \text{GeV}) / \text{m}^3$}
    \put(27,785){\includegraphics[width=1000\unitlength]{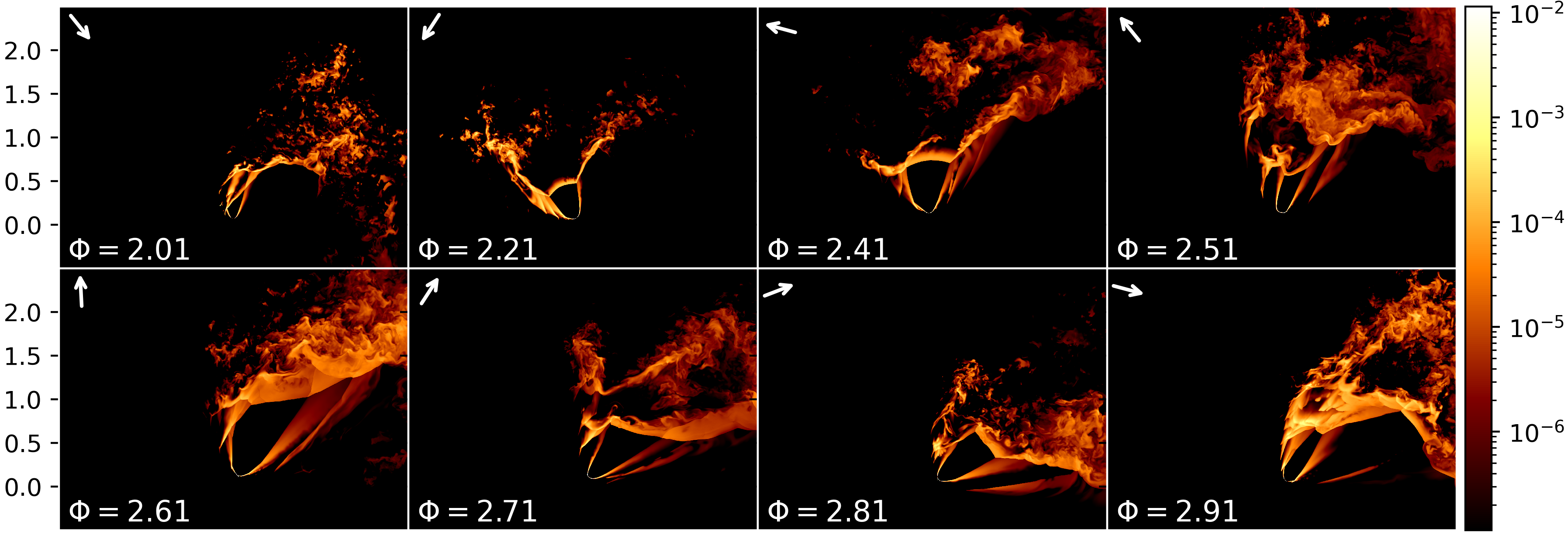}}

    \put(2,490){\rotatebox{90}{$y/\text{AU}$}}
    \put(2,650){\rotatebox{90}{$y/\text{AU}$}}
    \put(810,770){$n_l'(E = 485.5\ \text{GeV}) / \text{m}^3$}
    \put(27,416){\includegraphics[width=1000\unitlength]{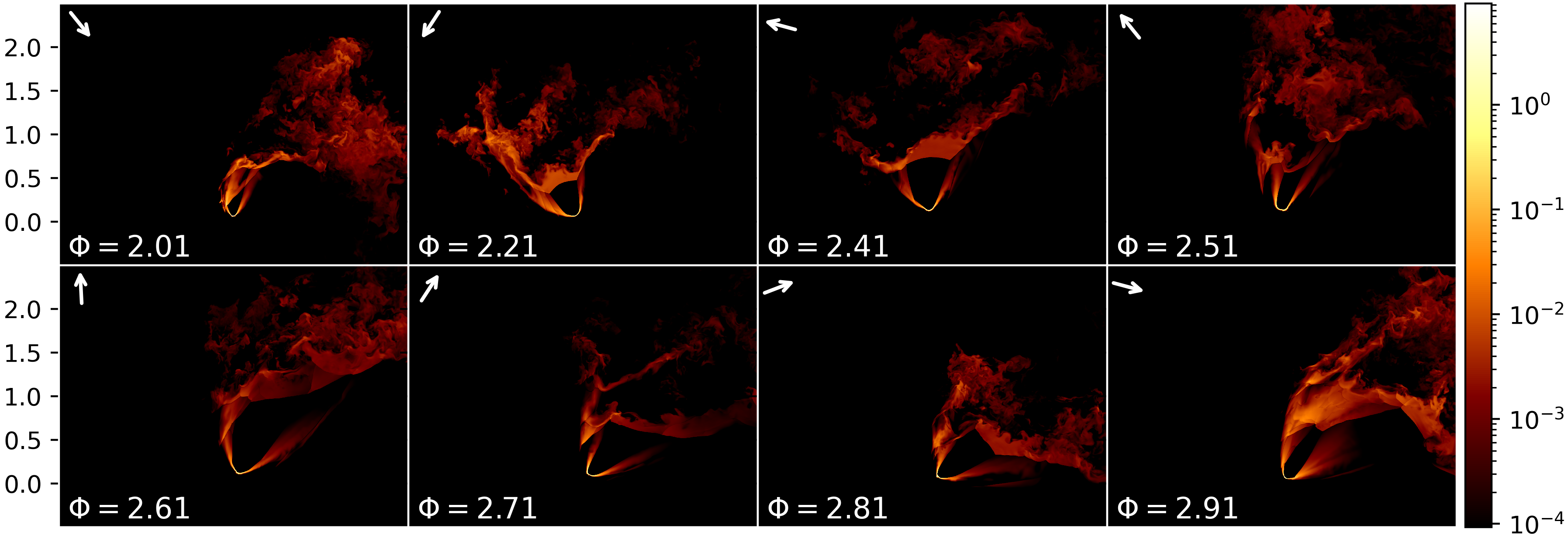}}

    \put(160,2){$x/\text{AU}$}
    \put(380,2){$x/\text{AU}$}
    \put(600,2){$x/\text{AU}$}
    \put(820,2){$x/\text{AU}$}
    \put(2,120){\rotatebox{90}{$y/\text{AU}$}}
    \put(2,290){\rotatebox{90}{$y/\text{AU}$}}
    \put(830,402){$n_l'(E = 1.5\ \text{GeV}) / \text{m}^3$}
    \put(27,29){\includegraphics[width=991\unitlength]{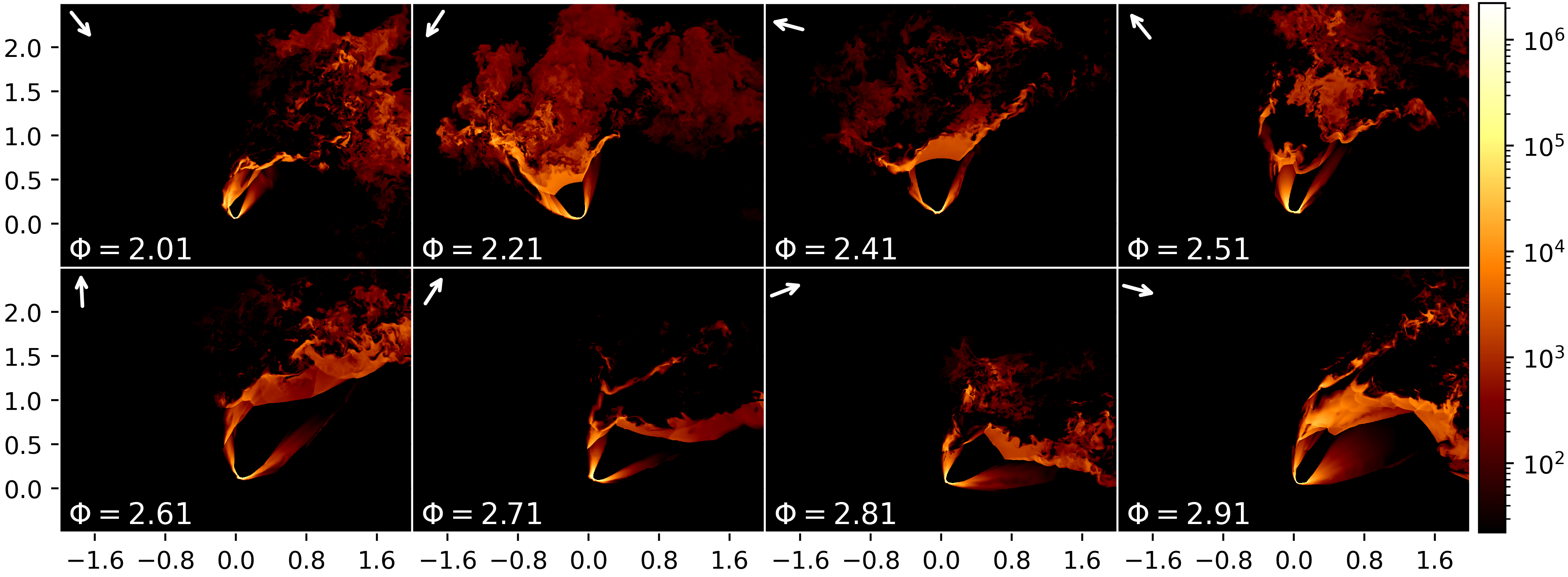}}

  \end{picture}
  \caption{\label{FigElectronsPlane}
        Number density of accelerated leptons in the orbital plane.
        Results for energies of 1.5\ GeV (bottom), 485.5\ GeV (middle), and 5467.7\ GeV (top) as indicated above the corresponding plots.  
        We show our results for eight different orbital phases of the second orbit as indicated in the plot.  
        In each case, we depict a range of densities from the peak value at periastron to a factor of $10^{-5}$ of this value.
        Additionally, the white arrows indicate the direction towards the observer.
        An example for corresponding fluid quantities is shown in Fig. \ref{FigCompQuantities} in Appendix \ref{AppendComp}.
  }
  
\end{figure*}
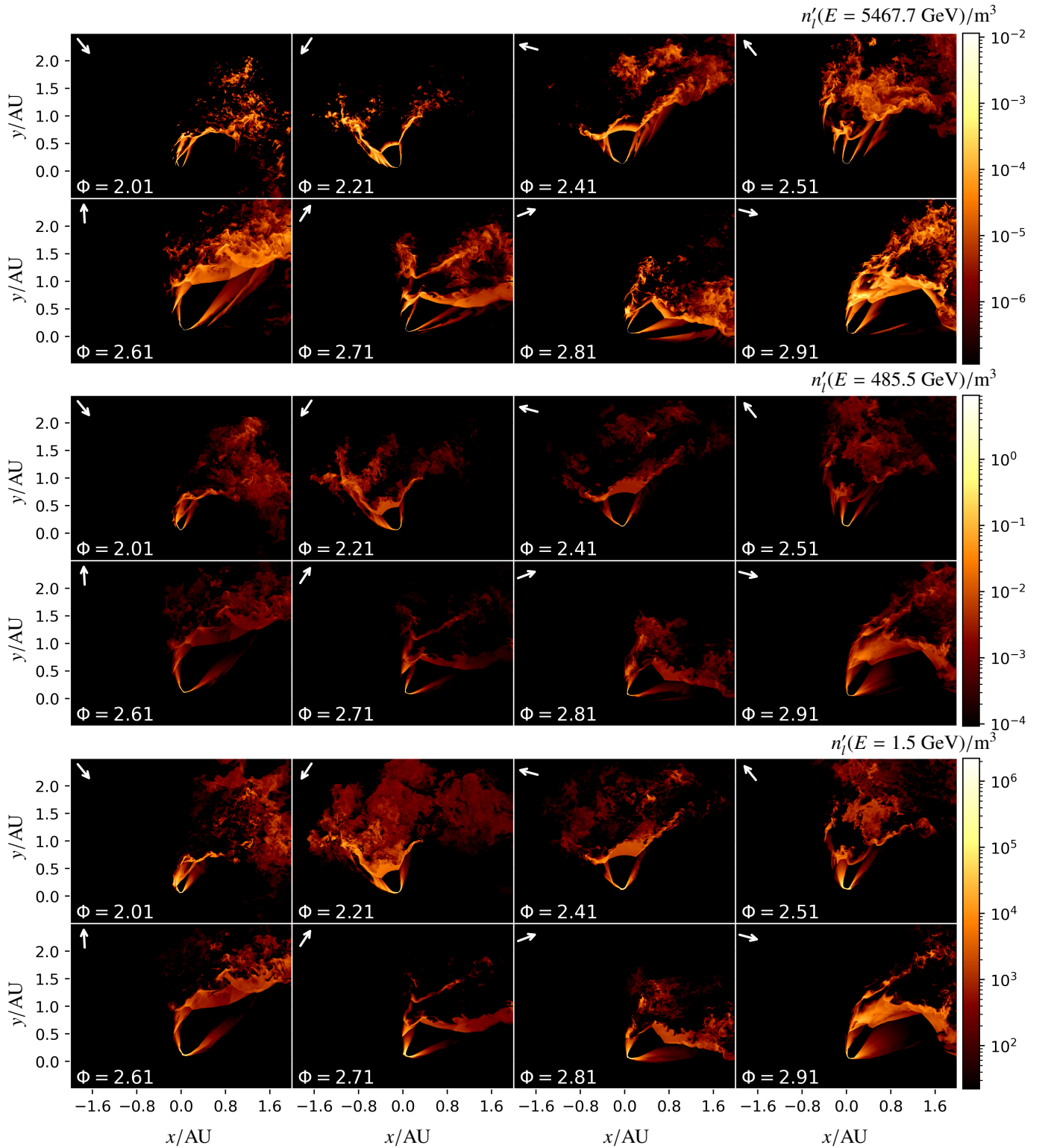

\begin{figure*}
  \setlength{\unitlength}{0.0175268cm}
  \begin{picture}(1027,1024)(0,0)

    \put(2,768){\rotatebox{90}{$z/\text{AU}$}}
    \put(2,908){\rotatebox{90}{$z/\text{AU}$}}
    \put(800,1016){$n_l'(E = 5467.7\ \text{GeV}) / \text{m}^3$}
    \put(27,705){\includegraphics[width=1000\unitlength]{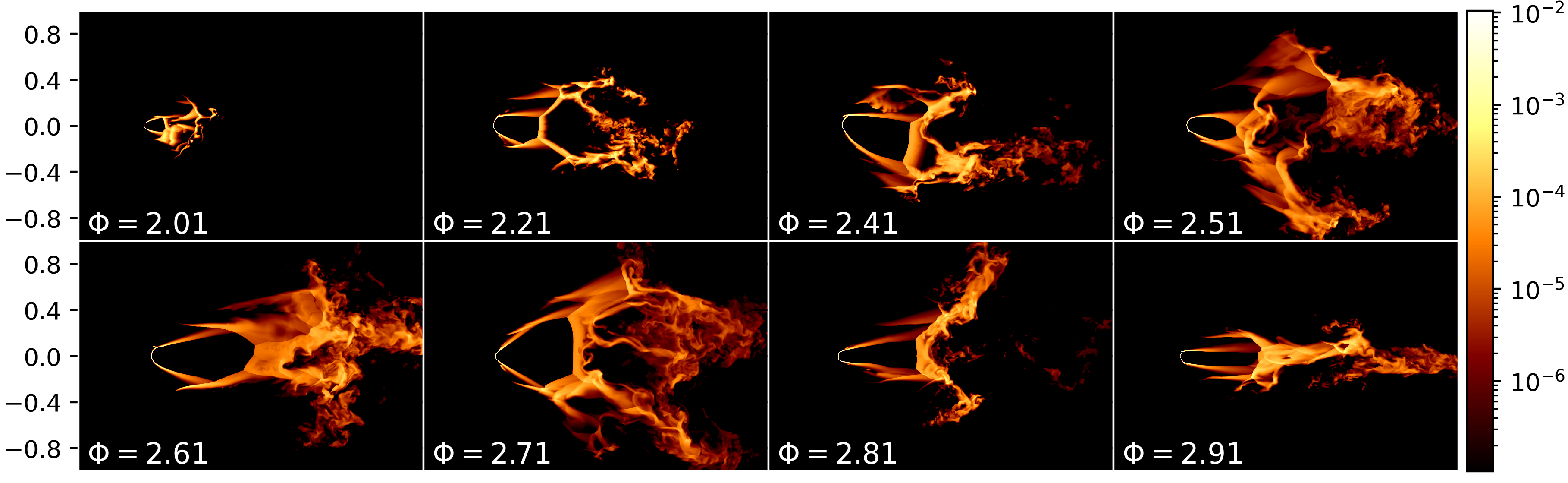}}

    \put(2,440){\rotatebox{90}{$z/\text{AU}$}}
    \put(2,580){\rotatebox{90}{$z/\text{AU}$}}
    \put(810,688){$n_l'(E = 485.5\ \text{GeV}) / \text{m}^3$}
    \put(27,377){\includegraphics[width=1000\unitlength]{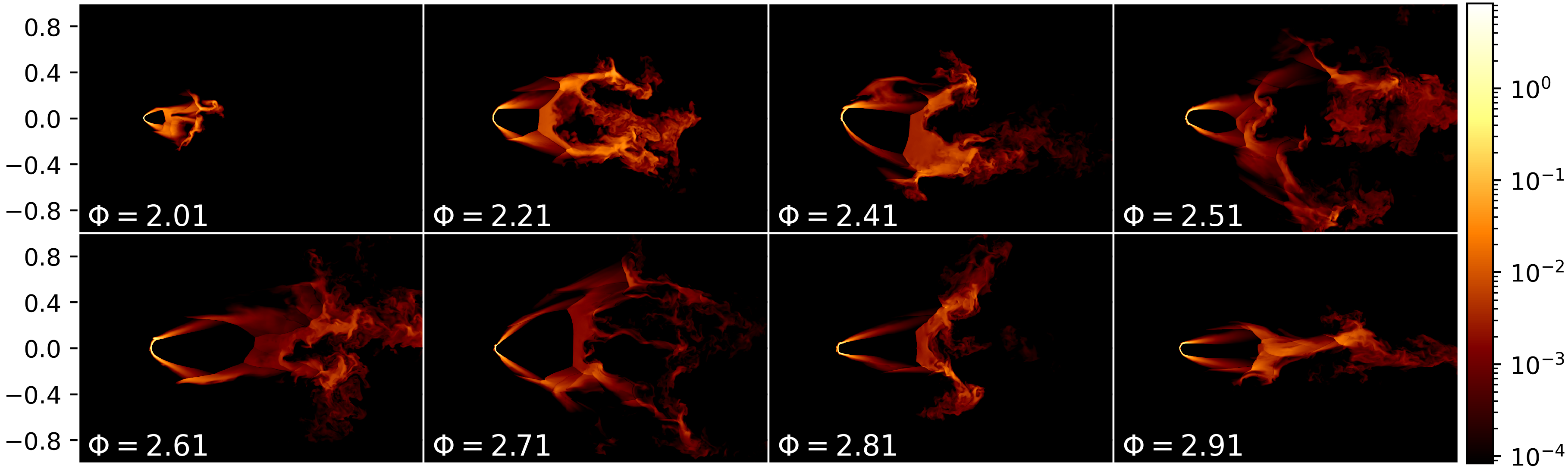}}

    \put(160,2){$x/\text{AU}$}
    \put(380,2){$x/\text{AU}$}
    \put(600,2){$x/\text{AU}$}
    \put(820,2){$x/\text{AU}$}
    \put(2,110){\rotatebox{90}{$z/\text{AU}$}}
    \put(2,250){\rotatebox{90}{$z/\text{AU}$}}
    \put(830,360){$n_l'(E = 1.5\ \text{GeV}) / \text{m}^3$}
    \put(27,29){\includegraphics[width=991\unitlength]{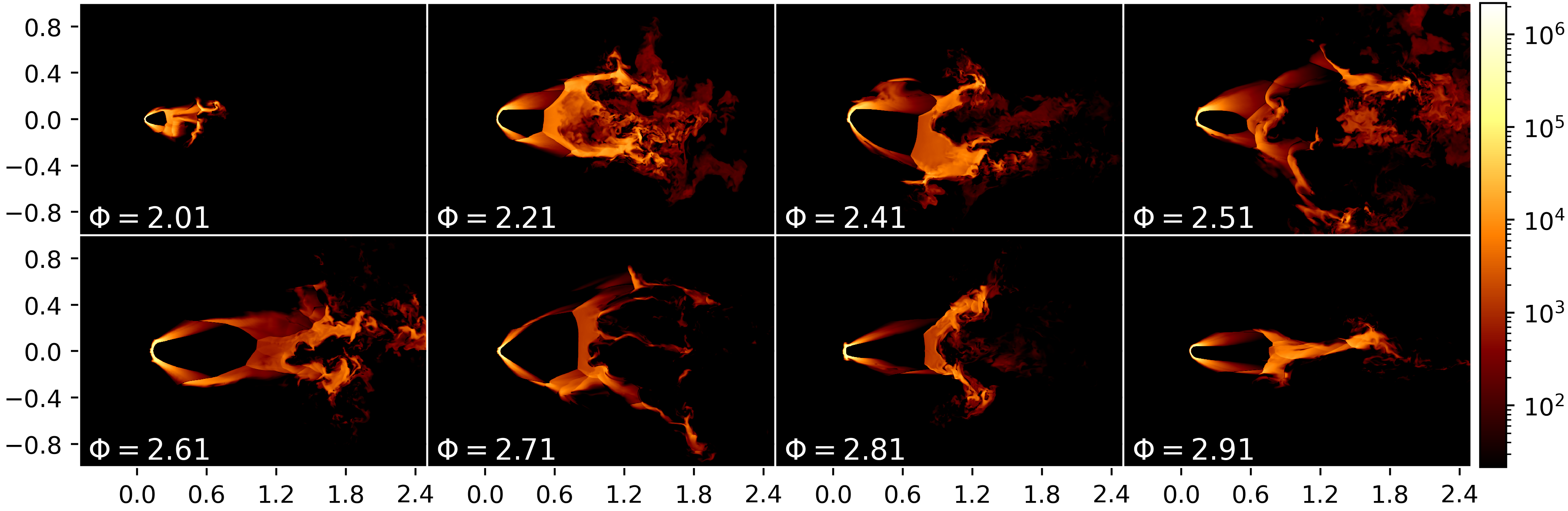}}

  \end{picture}
  \caption{\label{FigElectronsPerpendicular}
        Same as Fig.~\ref{FigElectronsPerpendicular}, but perpendicular to the orbital plane, where the plane displaying the results is oriented along the line connecting the star and the pulsar.
  }
  
\end{figure*}

The resulting distribution of energetic leptons for different phases and different particle energies in our simulation is shown in Figs. \ref{FigElectronsPlane} and \ref{FigElectronsPerpendicular}\footnote{
To relate these distributions to those of the wind plasma in the system, we show distributions for different plasma properties for a single phase in Fig. \ref{FigCompQuantities} in Appendix \ref{AppendComp}.}.
In both figures, we show the number density of leptons at energies of 1.5\ GeV, 485.5\ GeV, and 5.48\ TeV. 
In Fig.~\ref{FigElectronsPlane}, we show the lepton densities in the orbital plane, while Fig.~\ref{FigElectronsPerpendicular} gives the distribution for the same phases perpendicular to the orbital plane.
In the latter case, we show the data for the plane constructed from the line connecting both stellar objects and the $z$-axis.
We observed a pronounced asymmetry perpendicular to the orbital plane driven by the turbulent interaction of the winds.

In most cases (apart from some phases at the highest energy), the injection sites are the locally brightest regions.
Since the particles lose energy by adiabatic cooling through expansion and radiative losses when propagating away from the injection sites, the density of high-energy particles decreases, accordingly, away from the shocks.
Additionally, the injection rate is highest in close vicinity to the pulsar (i.e. at the nose of the WCR), since the wind density, by which particle injection is scaled, is highest there.
The ensuing non-thermal emission is discussed below, but from the figures it can be expected that the lower density of energetic leptons in the tail of the WCR will end up being partially compensated for by its greater volume.

When comparing the results at the different energies, the impact of the higher losses for increasing particle energy becomes apparent.
At an energy of $\sim5$~TeV at the bow shock of the WCR, the energetic leptons are strongly confined to this injection site.
This is caused by the strong magnetic field and the intense radiation field near the O-type star.
For the Coriolis shock, the distribution around the shock is more extended, where we observe spatial loss scales (seen in the slope in the energetic-lepton density in the downstream direction) of the order of 0.1~AU.
These loss scales clearly depend on the distance from the star, namely, they become shorter near periastron and for a smaller volume of the unshocked pulsar wind.
Finally, at $\sim5$~TeV we found a smaller contrast between the nose region and the tail region than at lower energies.
This indicates a suppression of the injection at these high energies in the nose region due to the high energy losses, there.
At the same time, these high energy-loss rates also indicate correspondingly high emission of non-thermal radiation.

At an energy of $\sim$1.5~GeV, we mainly observe leptons at the peak of the transported Maxwellian distribution.
In this case, the particles beyond the Coriolis shock are transported further in the downstream direction with their density hardly decreasing at all.
At the bow shock, they are still confined to the vicinity of the shock, but fill a much larger volume than at $\sim 5$~TeV.

At an energy of 485.5~GeV the situation is between both extreme cases.
Here, we visualise the distribution of particles within the power law regime, but far below its cut-off energy.
While the energy losses do not have a visual impact via obvious localised gradients, the particle distribution at the shocks in the nose region are markedly thinner than at $\sim$1.5~GeV.

In contrast to our previous model \citet{HuberEtAl2021AnA649_71}, the Coriolis shock is fully contained in the simulation at all phases.
As a result, we observe a broad distribution of energetic leptons downstream of the unshocked pulsar wind at all times.
These do not only originate from the Coriolis shock, but also from additional localised shocks in the turbulent downstream medium.
Correspondingly, we also expect some changes in the non-thermal emission with respect to the previous study related to the additional leptons in the downstream region.
This is discussed in the next section.

\begin{figure}
\resizebox{\hsize}{!}{
        \setlength{\unitlength}{0.0081cm}
        \begin{picture}(1040,691)
                \put(2,300){\small \rotatebox{90}{$\gamma^2 d N / d \gamma$}}
                \put(550,0){\small $\gamma$}
                \put(40,24){\includegraphics[width=1000\unitlength]{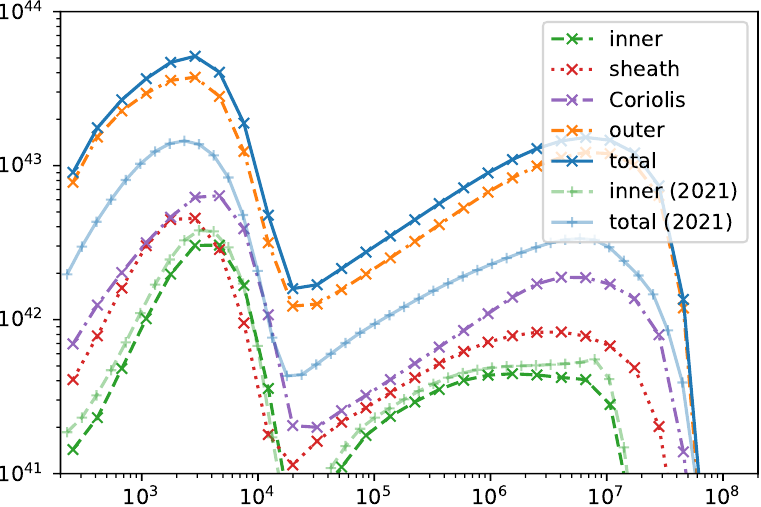}}
        \end{picture}
}
\caption{\label{FigParticleSpectrumContrained}
Domain-integrated spectral energy distribution of energetic leptons for different sub-regions, as discussed in the text.
Results are shown for an orbital phase $\phi=0.414$.
For comparison, we give the total spectral energy distribution and the one in the inner region from \citet{HuberEtAl2021AnA649_71} for a very similar orbital phase ($\phi = 0.393$) via the corresponding curves with a lower opacity. See label 2021 in the legend.
}
\end{figure}

As a further illustration of these differences of the particle content compared to our previous study, we show sub-domain integrated SEDs of the energetic leptons for an orbital phase $\phi=0.414$ shortly before apastron in Fig.~\ref{FigParticleSpectrumContrained}, where
\[
\frac{\text{d}N}{\text{d}\gamma} = \int_V n_l' \text{d}^3x'.
\]
The different sub-regions indicated in that figure are hollow spheres around the origin with different inner and outer radii.
The inner region with a radius of 0.21\ AU around the origin extends up to 1.5 times the distance of the bow shock from the origin along the line connecting star and pulsar in the direction of the pulsar.
In a similar fashion, the sheath extends from there up to a radius of 3/4 of the Coriolis shock distance, namely, up to 0.54\ AU.
Next, the Coriolis-shock region extents up to 5/4 of the Coriolis-shock distance or 0.91 AU, which is followed by the outer region, representing the rest of the numerical domain (for an illustration of these sub-regions, see Fig.~\ref{FigAnalysisRegionsParticleSpectra} in Appendix \ref{AppendAnalysisRegions}).
For comparison, we also show corresponding results from \citet{HuberEtAl2021AnA649_71} for a similar orbital phase $\phi = 0.393$, where we show the SED in the inner region (here with a radius of 0.22\ AU) and the total SED.

Apparently, the flux of energetic particles in the respective inner region is very similar in both cases. 
In contrast, it becomes once more obvious that in the present study the contribution of the outer region to the total particle flux is significantly higher.
This is a consequence of the much larger spatial domain considered in the  present simulation, which led to the acceleration site of the Coriolis shock being present at all times.

Additionally, the present simulation contains a much larger volume of highly turbulent downstream medium than the previous simulation, which would further increase in a model with an even larger domain.
However, Figs.~\ref{FigElectronsPlane} and \ref{FigElectronsPerpendicular} show that at $E=5$~TeV, the flux of non-thermal leptons directly downstream of the Coriolis shock is much higher than further in the turbulent downstream medium.
Thus, we expect the impact of this medium for the high-energy non-thermal particles to decrease further from the stars.
Additionally, shocks in this turbulent medium are weaker than the Coriolis shock \citep[][]{Bosch-RamonEtAl2012AnA544_59}.
Thus, an improved injection model taking into account the shock compression ratio \citep[][]{JonesEtAl1999ApJ512_105, VaidyaEtAl2018ApJ865_144V, WinnerEtAl2019MNRAS488_2235} can be expected to lead to an even smaller impact of the downstream medium.
Finally, while the increase in total flux compared to the previous model is around a factor of 5, we do not have to expect a similar increase in non-thermal radiation, because in the  outer regions non-thermal emission decreases due to the weaker magnetic field and stellar radiation field.
Thus, we expect that the present simulation covers the relevant parts of the system for the description of the non-thermal emission from the WCR.

\begin{figure}
\resizebox{\hsize}{!}{
        \setlength{\unitlength}{0.0081cm}
        \begin{picture}(1040,678)
                \put(2,300){\small \rotatebox{90}{$\gamma^2 d N / d \gamma$}}
                \put(550,0){\small $\gamma$}
                \put(40,24){\includegraphics[width=1000\unitlength]{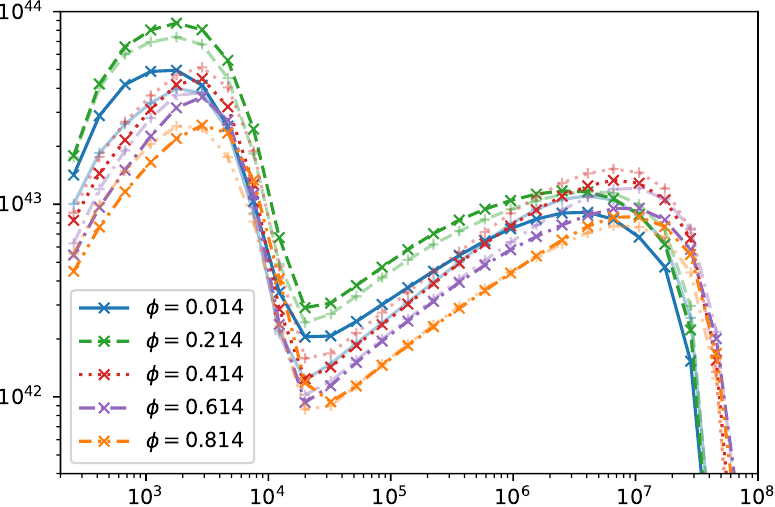}}
        \end{picture}
}
\caption{\label{FigParticleSpectraOrbits}
Evolution of the domain-integrated spectral energy distributions of the energetic particles for the second and the third (at a lower opacity and using \texttt{+} as a marker) orbit of the simulation.
Spectra are given as a function of the particle's Lorentz factor.
}
\end{figure}

In Fig.~\ref{FigParticleSpectraOrbits}, we show the particle SEDs integrated over the full numerical domain for different phases of the simulation.
In particular, we show data for both orbits contrasted against each other.
Interestingly, we do not find the largest orbit-to orbit variations after apastron, where we observed the strongest fluctuations in the wind properties \citep[see][]{KissmannEtAl2023AnA677_A5}.
Instead the orbit-to-orbit variation is largest around periastron in this case.

\begin{figure}
\resizebox{\hsize}{!}{
        \setlength{\unitlength}{0.0081cm}
        \begin{picture}(1050,668)
                \put(0,330){\small \rotatebox{90}{$E_\text{total}^{\text{CR}} / \text{J}$}}
                \put(340,0){\small $\phi$}
                \put(800,0){\small $\phi$}
                \put(50,50){ \includegraphics[width=1000\unitlength]{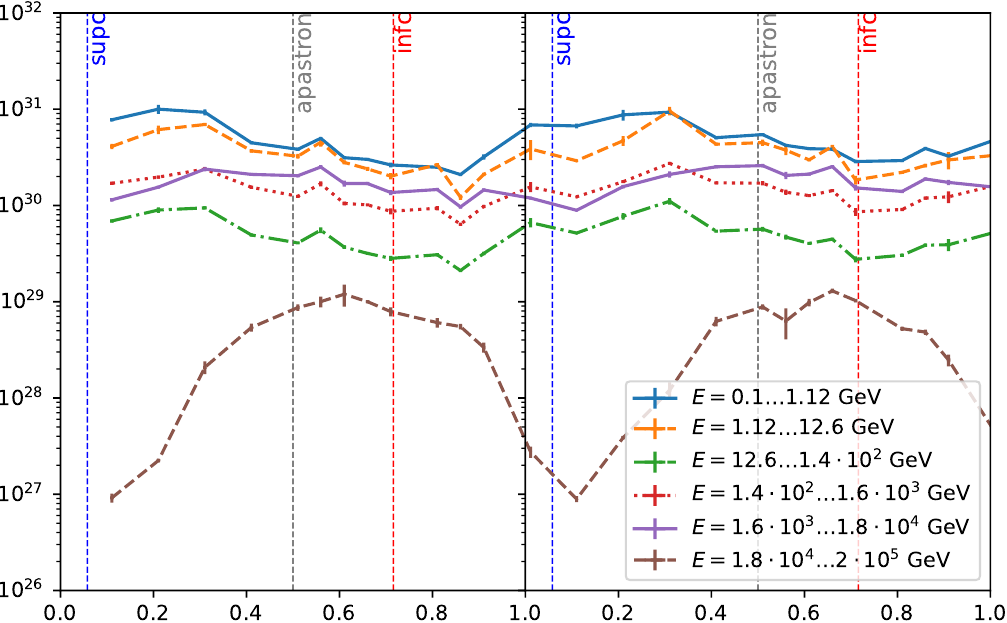} }
        \end{picture}
}
\caption{\label{FigIntegratedElectrons}
Temporal evolution of the integrated energy of energetic leptons $E_\text{total}^{\text{CR}}$ within the total computational domain for several bins of particle energy as indicated in the plot. 
We show the results for the second (left) and third (right) orbit of our simulation.
Relevant orbital phases are indicated in the plot via vertical lines, where additionally periastron is at $\phi=0$.
The error bars indicate the short-time variation in the six phases used to compute each data point.
}
\end{figure}

To further quantify the orbital dynamics, we show the orbital variation of the integrated energy of energetic leptons $E_\text{total}^{\text{CR}}$ in Fig.~\ref{FigIntegratedElectrons}.
Here, we integrated the lepton density  $n_l'$ for all orbital phases over the whole simulation domain and over those energy regimes as indicated in the figure,
\begin{equation}
E_\text{total}^{\text{CR}} = \int\limits_V \int\limits_{\gamma_\text{min}}^{\gamma_\text{max}} E_0 \gamma n_l' \text{d}\gamma \text{d}V
,\end{equation}
with $E_0$ as the rest energy of the leptons.

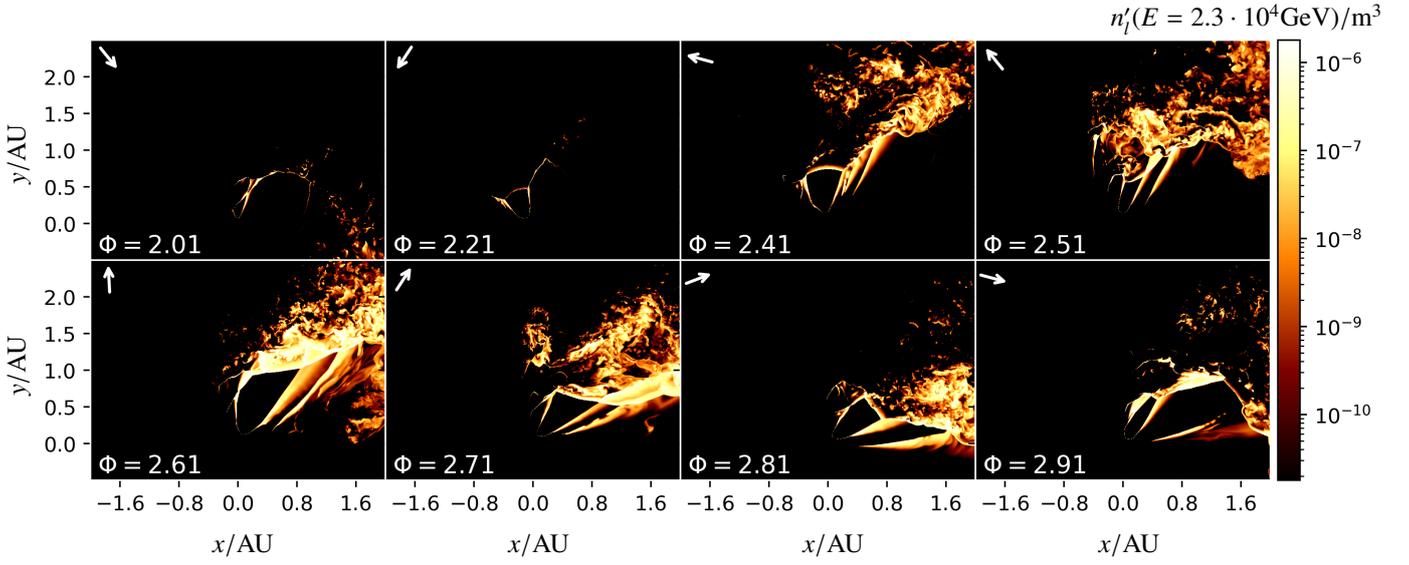
\begin{figure*}
\setlength{\unitlength}{0.0175268cm}
\begin{picture}(1027,410)(0,0)
\put(155,2){$x/\text{AU}$}
\put(377,2){$x/\text{AU}$}
\put(598,2){$x/\text{AU}$}
\put(820,2){$x/\text{AU}$}
\put(2,120){\rotatebox{90}{$y/\text{AU}$}}
\put(2,280){\rotatebox{90}{$y/\text{AU}$}}
\put(830,400){$n_l'(E = 2.3\cdot 10^4 \text{GeV}) / \text{m}^3$}
\put(27,29){\includegraphics[width=1000\unitlength]{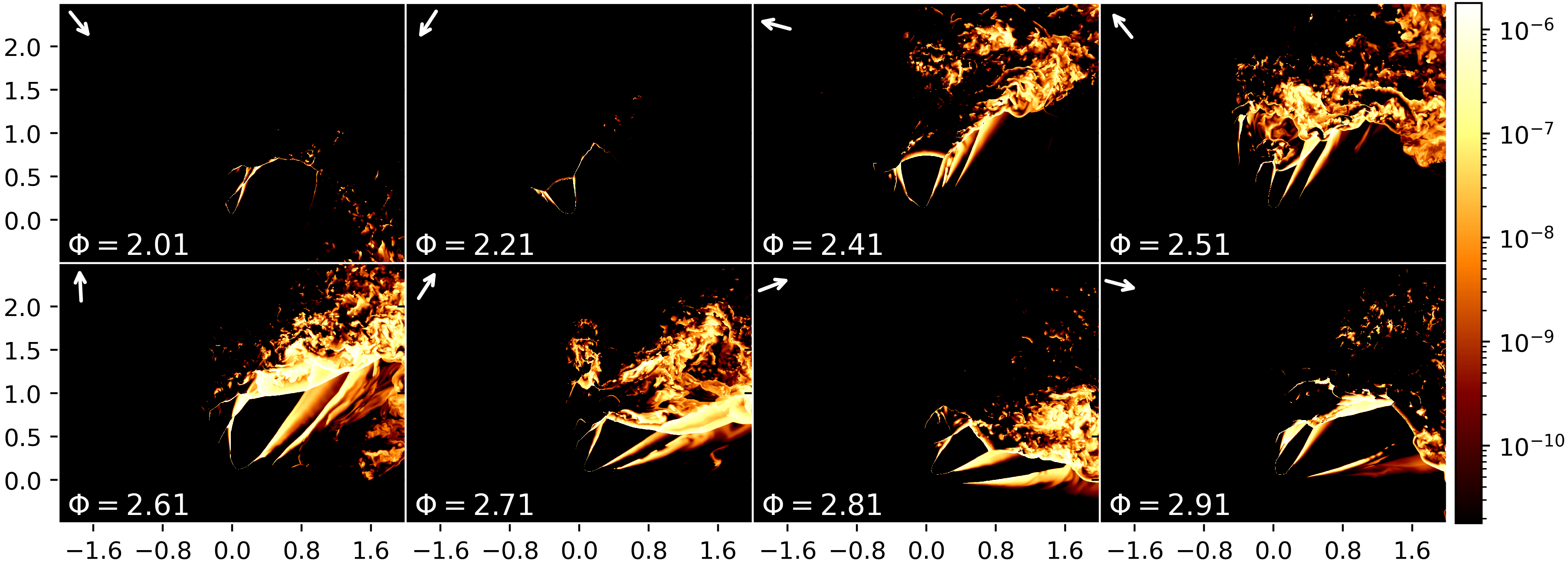}}
\end{picture}
\caption{\label{FigElectronsPlaneHighest} 
        Same as Fig.~\ref{FigElectronsPlane} but for an energy of $E=2.3\cdot10^4$~GeV, representative of the energy regime around the cut-off of the non-thermal lepton spectrum.
        }
\end{figure*}

Apparently, the peak of the particle distribution lags behind periastron, where we observe the highest flux at $\phi=0.214$ in Fig.~\ref{FigParticleSpectraOrbits}.
This was already found by \citet{HuberEtAl2021AnA649_71}, who discuss that this is is caused by the inertia of the fluid, at least for energies below $\sim$1 TeV.
For higher energies, however, we observe an even stronger delay: in the range of $1.6 \dots 18$~TeV we find a peak of the corresponding integrated particle energy near apastron.
The total energy of the highest energy leptons peaks even beyond that.
This is related to what we see in Fig.~\ref{FigElectronsPlaneHighest}, where we show the distribution of energetic leptons with an energy of $E=2.3\cdot10^4$~GeV, around the cut-off energy of the electron spectrum.
Here, we only find particles at a sufficiently large distance from the star.
At periastron, where the relevant acceleration sites are closest to the star, acceleration does not reach this energy any more.
Only around and particularly after apastron do we find a significant distribution of electrons at these high energies. 
According to Fig.~\ref{FigElectronsPlaneHighest}, a significant fraction of these particles are accelerated at the Coriolis shock, which was missing in the previous study especially for phases around and after apastron.
Additionally, we find a high density of these very-high-energy electrons also in the tail of the WCR, which are partly accelerated at localised shocks in the turbulent medium, and partly transported downstream from the large-scale shocks in the vicinity of the pulsar-wind cavity.

These stark differences as compared to the previous model also lead to the expectation of corresponding changes in the emission of non-thermal radiation.
As discussed above, the non-thermal emission cannot be expected to follow the distribution of energetic leptons due to the strong radial dependence of the corresponding target fields.
Emission from the simulated system is discussed in the next section.

\subsection{Non-thermal emission}
\label{SecEmissionNonMod}

\begin{figure*}
\centering
\setlength{\unitlength}{0.01698cm}

\begin{picture}(1040,683)(0,0)
\put(10,140){\small \rotatebox{90}{$\displaystyle E_{\gamma}^2 F_{E_\gamma} / \text{GeV s}^{-1} \text{m}^{-2}$}}
\put(10,450){\small \rotatebox{90}{$\displaystyle E_{\gamma}^2 F_{E_\gamma} / \text{GeV s}^{-1} \text{m}^{-2}$}}
\put(275,10){\small $\displaystyle E_{\gamma} / \text{GeV}$}
\put(750,10){\small $\displaystyle E_{\gamma} / \text{GeV}$}
\put(40,40){\includegraphics[width=1000\unitlength]{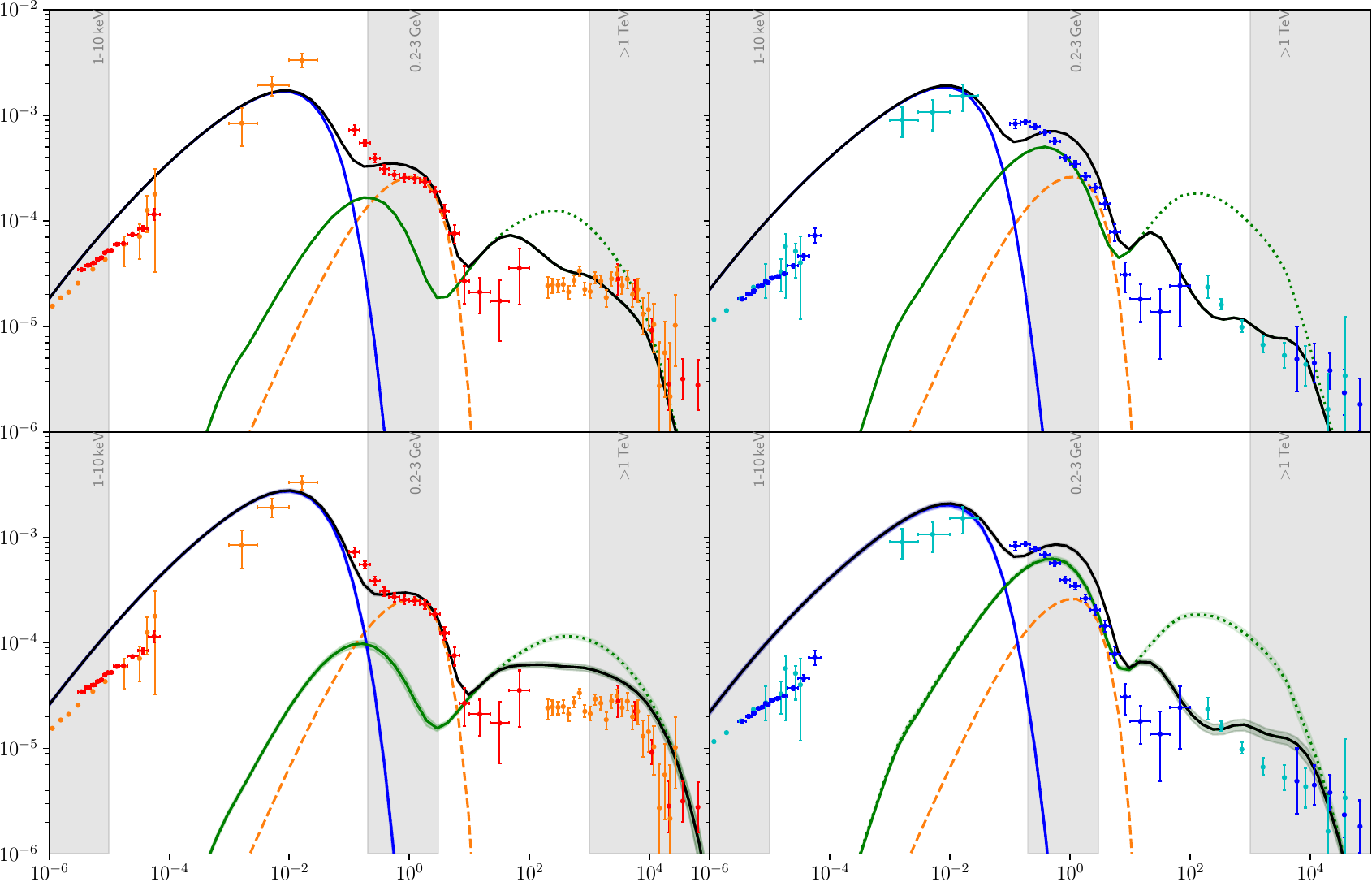}}
\put(500,320){\small \textcolor{red}{\textsc{Infc}}}
\put(980,320){\small \textcolor{blue}{\textsc{Supc}}}
\put(500,624){\small \textcolor{red}{\textsc{Infc}}}
\put(980,624){\small \textcolor{blue}{\textsc{Supc}}}
\put(122,320){\small $i=60^\circ$}
\put(605,320){\small $i=60^\circ$}
\put(122,624){\small $i=30^\circ$}
\put(605,624){\small $i=30^\circ$}

\end{picture}

\caption{Phase-averaged spectral energy distribution of the non-thermal emission predicted by our model for the \textsc{Infc} (left) and the \textsc{Supc} (right) period for  inclinations of $i=60^{\circ}$ (bottom) and $i=30^{\circ}$ (top).
The simulation results are shown together with observations in soft x-rays by Suzaku \citep[][orange and cyan]{TakahashiEtAl2009ApJ697_592} and NuSTAR \citep[][red and blue]{YonedaEtAl2021ApJ917_90}, as well as in LE gamma rays \citep[][orange and cyan]{CollmarZhang2014AnA565_38}, in HE gamma rays \citep[][red and blue]{YonedaEtAl2021ApJ917_90}, and in VHE gamma rays by H.E.S.S. \citep[][orange and cyan]{AharonianEtAl2006AnA460_743} and by HAWC \citep[][red and blue]{AlfaroEtAl2025ApJ987L_42}.
From the numerical model, we show synchrotron emission in blue, IC emission in green, and the magnetospheric emission of the pulsar \citep[][]{TakataEtAl2014ApJ790_18} via the orange dashed curve.
Additionally, we show IC emission without $\gamma\gamma$-absorption via the green dotted curve.
The standard deviation reflecting short-term variation of the spectrum is indicated by the transparent bands for the example of $i=60^\circ$.
The total emission is given by the black solid curve.
Additionally, we indicated the energies, over which we average to investigate the light curve in Fig.~\ref{FigLightCurvesFullAmplitude} via the grey bands.}
\label{FigSpectraINFCandSUPC}
\end{figure*}

From the distribution of HE particles, we computed the predicted non-thermal emission from the system for photon energies above 1~keV.
Corresponding results for the phase-averaged spectral energy distribution (SED) of the emission during the \textsc{Infc} and \textsc{Supc} phase intervals  are shown in Fig.~\ref{FigSpectraINFCandSUPC} together with observations in different energy regimes.
Here, we use the  definition for these phase intervals introduced by \citet{AharonianEtAl2006AnA460_743}; namely, a division of the orbit into the part $0.45 < \phi < 0.9$ (\textsc{Infc}) around inferior conjunction  and the rest (\textsc{Supc}) encompassing superior conjunction.
We refer to Fig. 1 of  \citet{HuberEtAl2021AnA649_71} or Fig. 4 of \citet{AharonianEtAl2006AnA460_743} for an illustration of the orbital phases.
Both our previous model \citet{HuberEtAl2021AnA649_71} and a first analysis of the present simulation showed that the emission of the energetic particles in the interacting winds (especially in the \textsc{Infc} phase interval) fall significantly below the observed HE gamma-ray emission.
These observations \citep[][]{YonedaEtAl2021ApJ917_90} hint at the presence of two emission components in this energy regime, where our previous model included one component in the form of the IC emission of the transported Maxwellian.
Considering that the emission from our simulated energetic leptons only represents the emission in the shocked pulsar wind, we added a model for the magnetospheric emission of the pulsar.
For this, we used the outer-gap model by \citet{TakataEtAl2014ApJ790_18} based on \citet{WangEtAl2010ApJ720_178}.
This model is independent of the pulsar phase and is therefore added as a time-independent component.
This is also consistent with the observed flux between 1 and 10 GeV hardly differing between the \textsc{Infc} and the \textsc{Supc} periods.
With the added magnetospheric emission, our model can nicely reproduce the cut-off in the SED at a few GeV.

From low to high photon energies, the dominating components of the SED are a power law of synchrotron emission up to a few MeV, followed by IC emission of the transported Maxwellian peaking at a few hundred MeV.
Next, we see the bump by the magnetospheric emission to be followed by the IC emission of the non-thermal leptons.
For the latter, there is a clear reduction of the radiative flux due to $\gamma\gamma$-absorption especially for the \textsc{Supc} phase interval.

Compared to our previous study \citep[][]{HuberEtAl2021AnA649_71} we find a higher flux in general, especially for an inclination of $i=60^\circ$.
Consequently, the model overestimates the flux in the synchrotron regime and from the upper end of the HE regime to the VHE regime.
Additionally, just below 100 GeV the model overshoots the HE data for both inclinations and both periods.
However, the spectral shape above $\sim$~10 GeV seems to be better represented by an inclination of $i=60^\circ$, where observations show a harder spectral index than we can reproduce with $i=30^\circ$.

In the \textsc{Supc} period, the spectral shape from the model around 1 GeV does not accurately represent the data.
This is the energy range, where the data indicate two emission components, which could be represented by the magnetospheric model and the transported Maxwellian.
With the higher energetic of these two components showing hardly any variation (see discussion above), we identify it with the magnetospheric emission.
With the lower energy component identified as the transported Maxwellian in our model, it actually peaks at an energy of about an order of magnitude too high when compared with observations.
A Maxwellian at such a lower energy and with a  correspondingly larger amplitude  can be expected to produce a more accurate match to the data.

Since we used the same injection parameters for the energetic leptons, we find a similar slope and extent for the synchrotron component as in our previous study \citet{HuberEtAl2021AnA649_71}, where we also over predicted the emission in this energy regime, especially in the \textsc{Supc} phase interval.
Especially compared to the NuSTAR data \citep[][]{YonedaEtAl2021ApJ917_90} it seems that our predicted synchrotron spectrum is harder by approximately 0.25.
However, a softer synchrotron spectrum would also imply a softer lepton spectrum and, thus, also a softer IC spectrum, which could produce discrepancies in the VHE regime.

Apparently, the NuSTAR cannot be smoothly extrapolated to the COMPTEL MeV data \citep[][]{YonedaEtAl2021ApJ917_90}.
The authors discuss possible reasons for this discrepancy, including a localised emission region with strong magnetic field \citep[][]{DubusEtAl2015AnA581_27} or IC emission of a population of lower energy electrons.
Both these scenarios can be tested with a future enhancement of our model.
Given that our present simulations are still based on relativistic hydrodynamics with just a crude approximation of the magnetic field without any directional information, we do not expect a good fit in this regime.
However, a future model using a dynamical magnetic-field model for the stellar winds to allow for a more accurate description of both synchrotron energy losses and synchrotron emission  might also produce such a highly-magnetised emission site.
Alternatively, a shifted Maxwellian in the lepton population might also help explaining the apparent peak in the MeV regime.

\begin{figure*} 
\setlength{\unitlength}{0.01747cm}
\begin{picture}(1030,350)(0,0)
\put(160,10){\small $x/\text{mas}$}
\put(402,10){\small $x/\text{mas}$}
\put(643,10){\small $x/\text{mas}$}
\put(885,10){\small $x/\text{mas}$}
\put(10,140){\small \rotatebox{90}{$y/\text{mas}$}}
\put(390,340){\small non-thermal energy flux $E^2_\gamma \Phi_{E_\gamma} / \text{GeV}\, \text{s}^{-1}\, \text{mas}^{-2}\, \text{m}^{-2}$}
\put(80,310){\small $E=31.6\text{keV}$}
\put(237,310){\small $E=1\text{MeV}$}
\put(394,310){\small $E=31.6\text{MeV}$}

\put(562,310){\small $E=15\text{GeV}$}
\put(719,310){\small $E=474\text{GeV}$}
\put(876,310){\small $E=15\text{TeV}$}
\put(27,30){\includegraphics[width=520\unitlength]{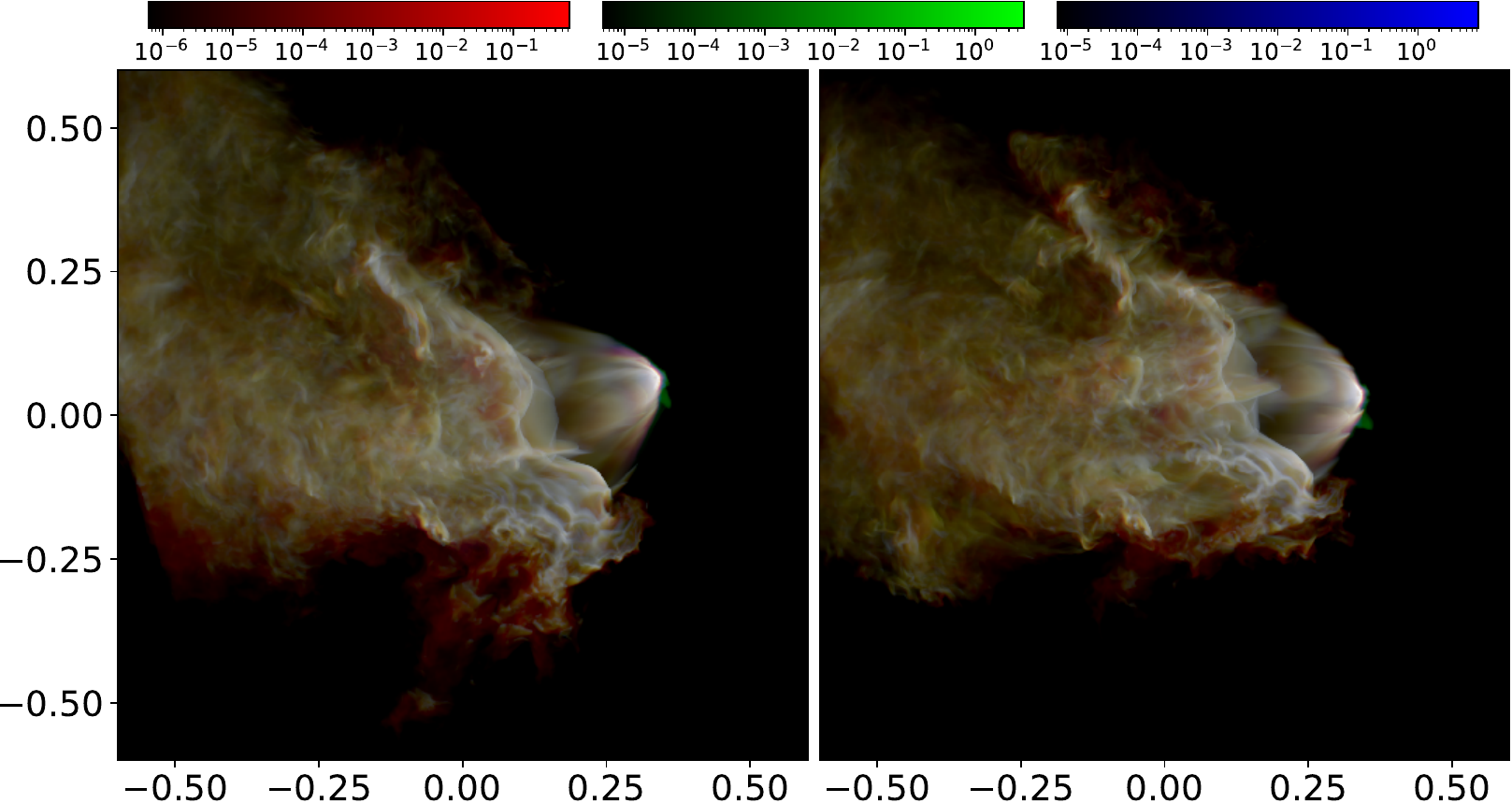}}
\put(550,30){\includegraphics[width=480\unitlength]{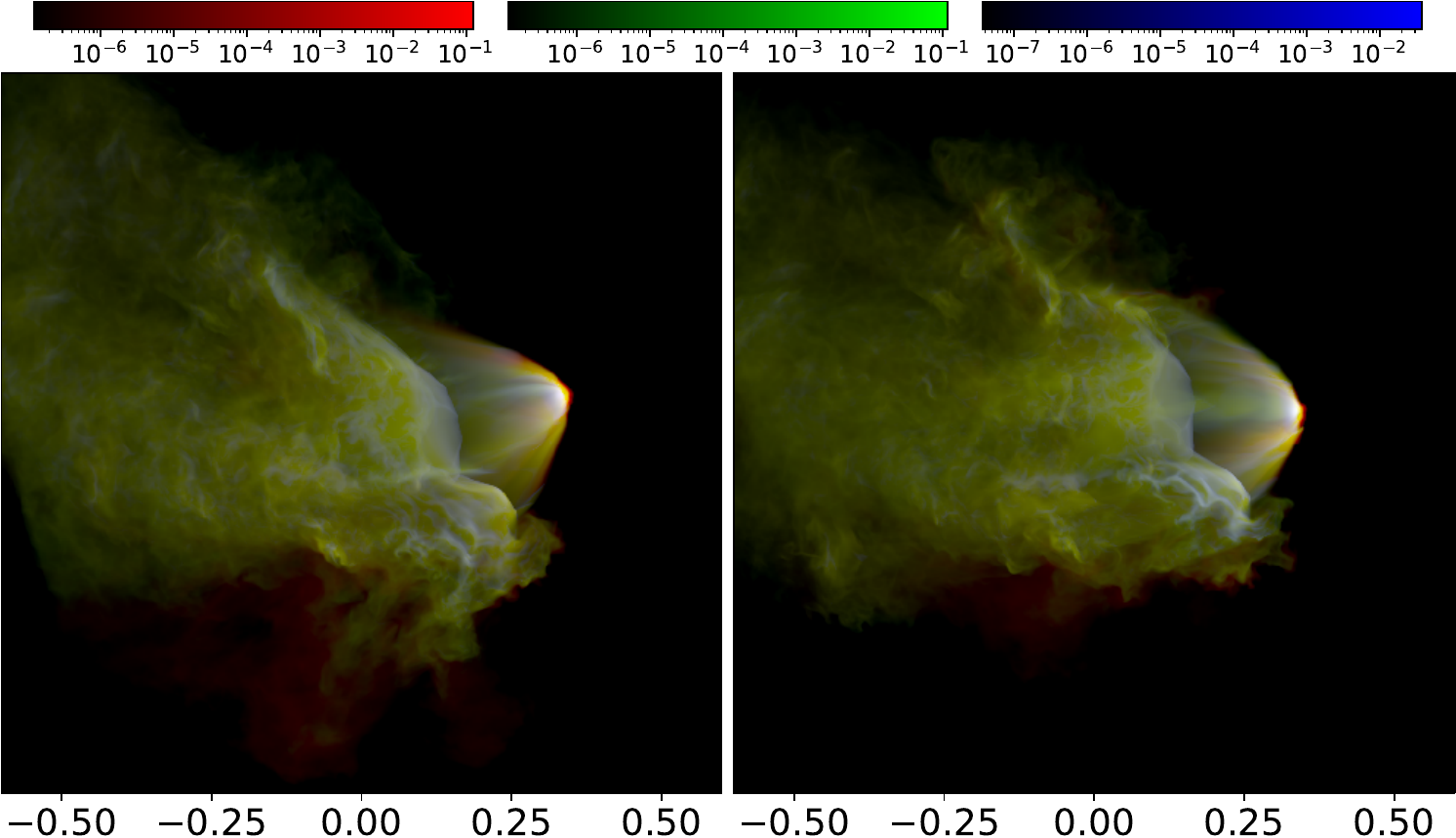}}

\put(250,260){\small\textcolor{white}{$i=30^\circ$}}
\put(490,260){\small\textcolor{white}{$i=60^\circ$}}
\put(730,260){\small\textcolor{white}{$i=30^\circ$}}
\put(970,260){\small\textcolor{white}{$i=60^\circ$}}

\end{picture}
\caption{\label{FigProjections}
Example projections of the predicted emission for the synchrotron regime (two left-handed plots) and the IC regime (two right-handed plots).
In both cases we show results for two different inclinations with $i=30^\circ$ on the left and $i=60^\circ$ on the right, respectively.
Results are shown for an orbital phase of $\phi=0.414$ in the third full orbit (see also Fig.~\ref{FigParticleSpectrumContrained}).
}
\end{figure*}
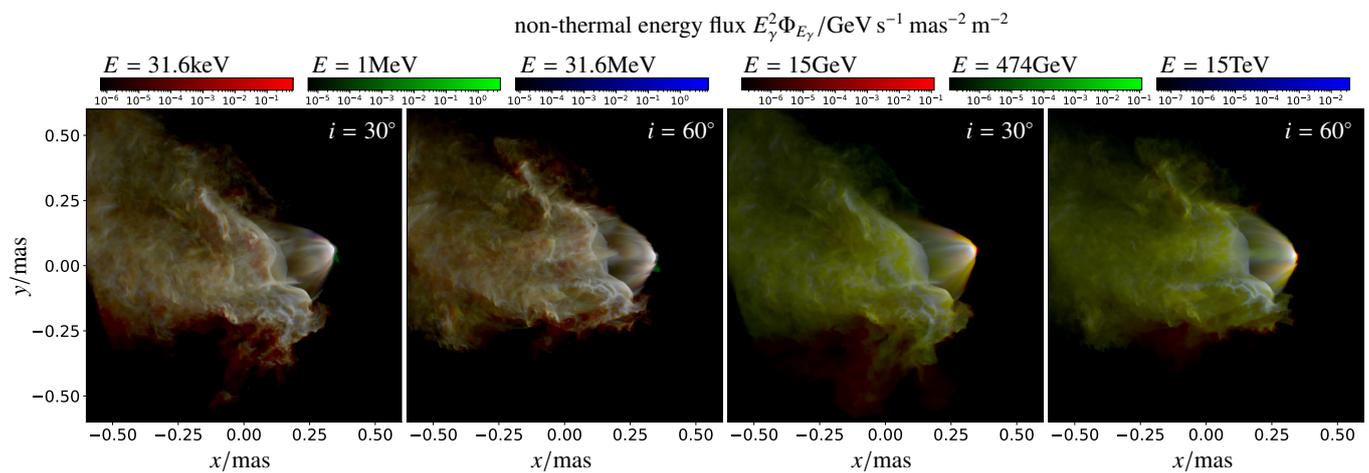

As an additional illustration we show false-color emission maps as they would be detectable from Earth with infinite angular solution in Fig.~\ref{FigProjections}.
Results are shown for the example of an orbital phase $\Phi=0.414$ in the third full orbit.
Here, we show emission maps for two different energy regimes and for the two inclinations investigated in this study. 
For the synchrotron regime, we use energies of 31.6~keV, 1~MeV, and 31.6~MeV, where at 31.6~MeV we have the beginning of the cut-off of the synchrotron spectrum and possibly some impact of the IC emission of the transported Maxwellian.
In the IC regime, the two lower energies are representative for the emission from the power-law of non-thermal electrons, whereas $E=15$~TeV again reflect the regime of the cut-off.

The different emission structures are apparent from Fig.~\ref{FigProjections}.
We see strong, localised emission at the bow shock.
Next to it we find emission from the sheath around the pulsar wind up to the Coriolis shock.
Beyond this, we see are large region with weaker emission from the turbulent tail of the WCR.
In the vicinity of the Coriolis shock, we see some additional shock structures that appear white in both energy bands, which indicates that these sites emit up to the highest energies.
In total it is apparent that the simulation domain contains the relevant emission sites.

Quantitatively the emission from the bow shock dominates the total emission from the system at all depicted energies.
In the synchrotron regime, this contributes slightly more than 50\% of the total emission at an energy of 1 MeV, whereas in the IC regime at 474 GeV the emission around the bow shock amounts to some 70\% of the total emission.
This becomes even more pronounced at higher energies (e.g. at 15~TeV the emission from the bow shock contributes 90\% of the total emission).
In contrast, the tail region beyond the Coriolis shock contributes approximately 26, 25, and 15\% of the total emission at the increasing energies of the synchrotron regime, with the rest stemming from the sheath region around the pulsar wind.
These values also show some inclination dependence, as can be seen from the additional results given in Appendix \ref{AppendQuantContributions}.
Additionally, we show an example for the definitions of the bow-shock and the sheath regions in Fig.~\ref{FigProjectionsMasks} in Appendix. \ref{AppendAnalysisRegions}.

This shows that despite the high flux of non-thermal particles in the tail of the WCR its contribution to the total emission of the system is rather small.
In the case of IC emission this is mostly due to the radially decreasing energy density of the stellar radiation field.
For the synchrotron regime the energy density of the target field -- in this case the local magnetic field -- is coupled to the energy density of the gas instead. 
But also for a more realistic magnetic field model, its decrease in the tail can be expected to be weaker than for the radiation field due to plasma dynamics, impact of shocks, and the relation to the pulsar magnetic field.
This also leads to the somewhat more pronounced structure visible in the synchrotron regime.

\begin{figure*}
\centering
\setlength{\unitlength}{0.01698cm}

\begin{picture}(1060,670)(0,0)
\put(310,660){\small{$i=30^\circ$}}
\put(800,660){\small{$i=60^\circ$}}
\put(10,110){\small \rotatebox{90}{$\displaystyle \frac{F_{> \text{1 TeV}}}{10^{-8} \text{m}^{-2} \text{s}^{-1}}$}}
\put(10,300){\small \rotatebox{90}{$\displaystyle \frac{F_{\text{0.2 - 3 GeV}}}{10^{-3} \text{m}^{-2} \text{s}^{-1}}$}}
\put(10,490){\small \rotatebox{90}{$\displaystyle \frac{F_{\text{3 - 10 keV}}}{10^{-14} \text{J\,m}^{-2} \text{s}^{-1}}$}}
\put(310,10){\small $\phi_\text{orbit}$}
\put(800,10){\small $\phi_\text{orbit}$}
\put(60,40){\includegraphics[width=1000\unitlength]{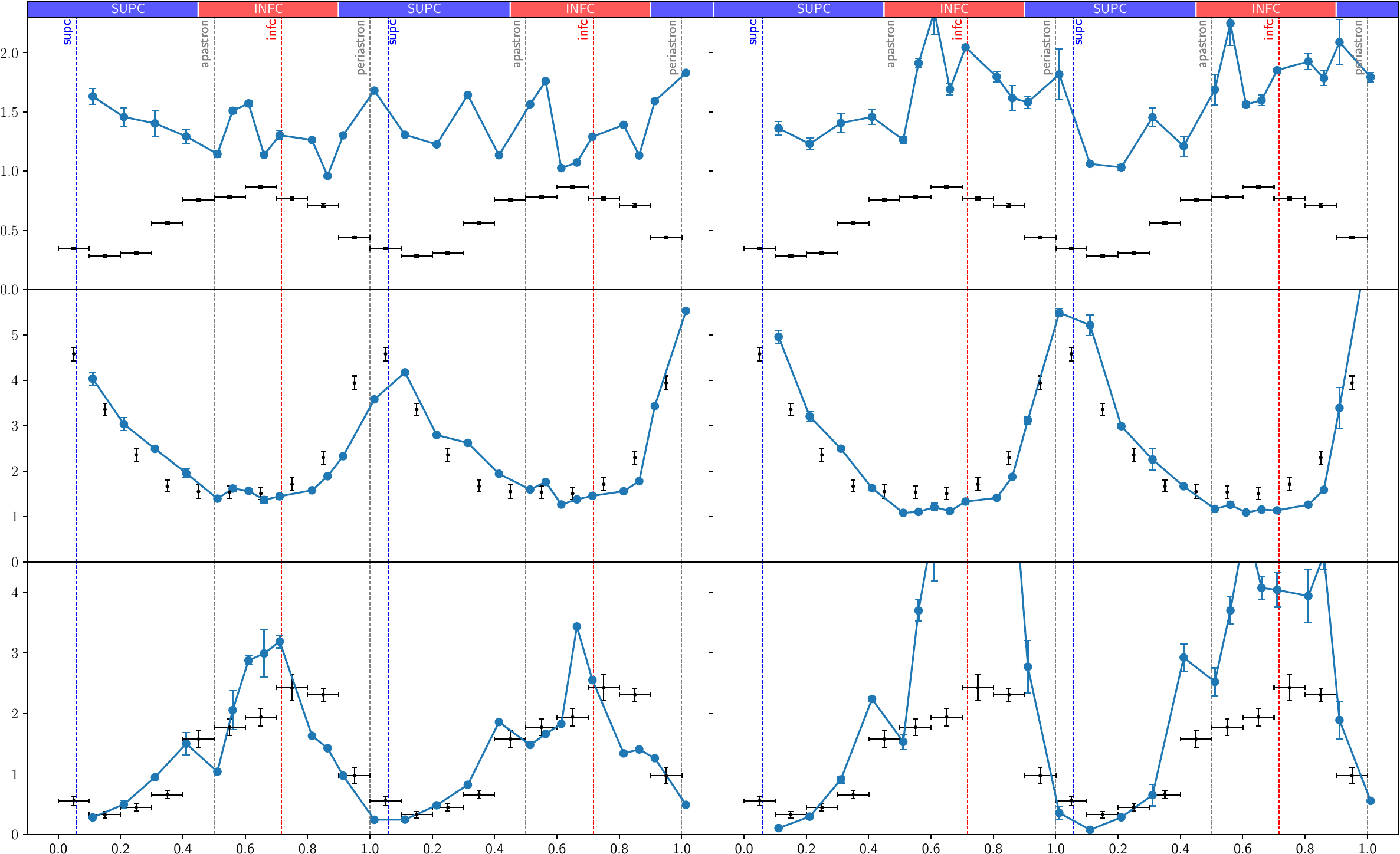}}
\end{picture}
\caption{\label{FigLightCurvesFullAmplitude}
Light curves of LS 5039 in three different energy bands and for two different inclinations ($i=30^{\circ}$ on the left and $i=60^{\circ}$ on the right).
Here, we show the model results in blue and corresponding data in black together with error bars where available.
The individual energy bands are soft x-rays at an energy of 3-10\,keV (top row, with data from \citet{YonedaEtAl2020PhRvL125_1103}, HE gamma rays at an energy of 0.2-3\,GeV (middle row, with data from \citet{HadaschEtAl2012ApJ749_54}), and VHE gamma-rays with energies above 1\,TeV (lower row, with data from \citet{AharonianEtAl2006AnA460_743}).}
\end{figure*}
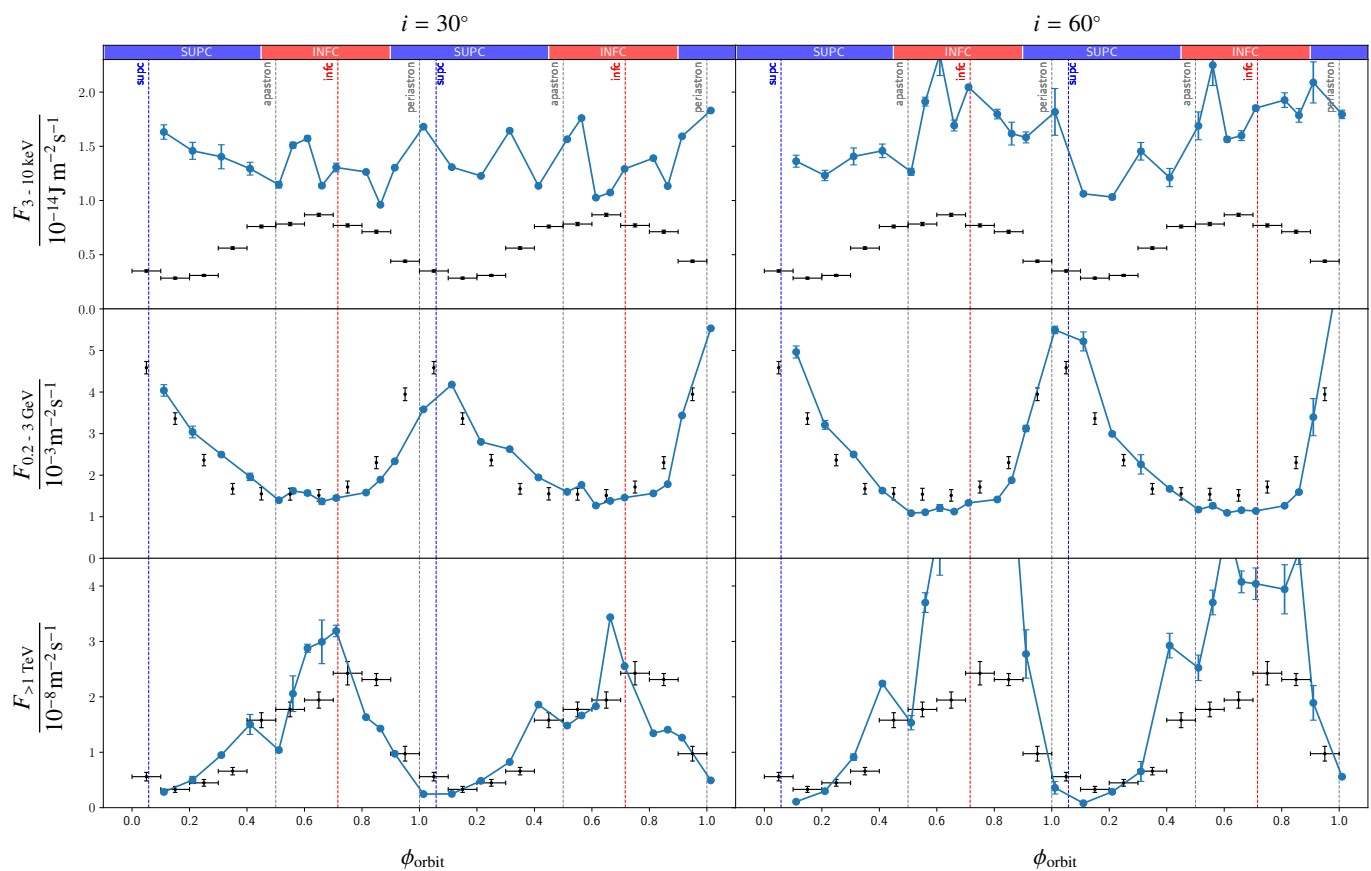

In Fig.~\ref{FigLightCurvesFullAmplitude} we show the corresponding non-thermal emission light curves at different energies for LS 5039, according to our numerical model for both inclinations compared to corresponding observational data.
Where available, the error bars of our simulation results represent the variation of the corresponding flux over a period of approximately 40\,min homogeneously covered by six individual output steps.
This variation does not necessarily represent stochastic variation by turbulence, but can also be related to a systematic effect, where $\gamma\gamma$-absorption quickly increases over a short timescale, when the relevant emission zone starts moving behind the star.

The light curves show some interesting changes compared to our previous study.
First, our prediction is much nearer to the observations in the HE gamma-ray regime, especially during the \textsc{Infc} phase interval.
This is simply due to the additional consideration of the magnetospheric emission via the model of \citet{TakataEtAl2014ApJ790_18,WangEtAl2010ApJ720_178}.
The remaining underflow as compared to the observations can be explained by the deficiencies of our model at the lower end of the HE gamma-ray regime, related to the emission of the transported Maxwellian.

As in our previous analysis, we find a significant overestimation in our model at keV energies for both inclinations, which was already apparent in Fig.~\ref{FigSpectraINFCandSUPC}.
Interestingly, for an inclination of $i=60^\circ$ the general dependence on orbital phase in this energy regime becomes more similar to observations than in our previous model.
In contrast, our previous model clearly featured the wrong phase dependence with peaks near superior conjunction.
At $i=30^\circ$, our new model does not seem to show a clear phase dependence at all.

The phase dependence at VHE gamma-ray energies is similar to our previous model for $i=30^\circ$.
However, for the case of the higher inclination angle, our new results become quite different from our previous model results for large parts of the \textsc{Infc} period.
This is connected to the missing particle population from the Coriolis shock in the previous model and to apparent differences in relativistic boosting.
The impact of the change in the particle population can be seen when comparing the emission light curve between the two models.
Here we observe an enhancement after apastron for both inclinations and both for synchrotron and VHE emission.

Regarding relativistic boosting, the previous model featured a significant peak in emission for $i=60^\circ$ around $\phi\simeq0.55$ from relativistic boosting in the leading edge in the WCR.
In contrast, we find significantly enhanced emission from $\phi\simeq0.55$ up to $\phi\simeq0.9$ in both investigated orbits of the new model, when comparing the $i=60^\circ$ results to those using $i=30^\circ$.
This enhancement is also visible in the synchrotron regime (top panels of Fig.~\ref{FigLightCurvesFullAmplitude}), where relativistic boosting is the only process that can change the amplitude of the emission, given that we are using a magnetic-field model that does not feature direction information, thus, leading to isotropic emission in the co-moving frame.
Here, the higher turbulence in the present model leads to relativistic flow in different directions, thus enhancing the range of orbital phases, for which relativistic boosting can lead to relevant changes in the flux of non-thermal emission.

\subsection{Non-thermal emission with modified injection}
Motivated by the observation that the model with the higher inclination angle represents a better match to observations in several aspects (spectral slope in the VHE regime for the \textsc{Infc} phase interval and phase-dependence of emission in x-rays) it seems that part of the reason for the deviation of this model from corresponding data might be related to an overestimate of the injection rate of leptons into the non-thermal power law.
Here, we have to take into account that we used the values from our previous study, because tuning for the injection parameters was impossible at the given high resolution.
However, the missing Coriolis shock in the previous model for phases around and after apastron apparently leads to a reduced emission in these phases.
This is especially relevant for leptons at the highest energies, as can already be expected from Fig.~\ref{FigElectronsPlaneHighest}, where we find the highest densities of VHE particles just at the orbital phases after apastron.

\begin{figure}
\resizebox{\hsize}{!}{
        \setlength{\unitlength}{0.0081cm}
        \begin{picture}(1050,792)(-50,-50)
                \put(-50,260){\small \rotatebox{90}{$E_{\gamma}^2 F_{E_\gamma}$ / GeV s$^{-1}$ m$^{-2}$}}
                \put(450,-50){\small $E_{\gamma}$ / GeV}
                \includegraphics[width=1000\unitlength]{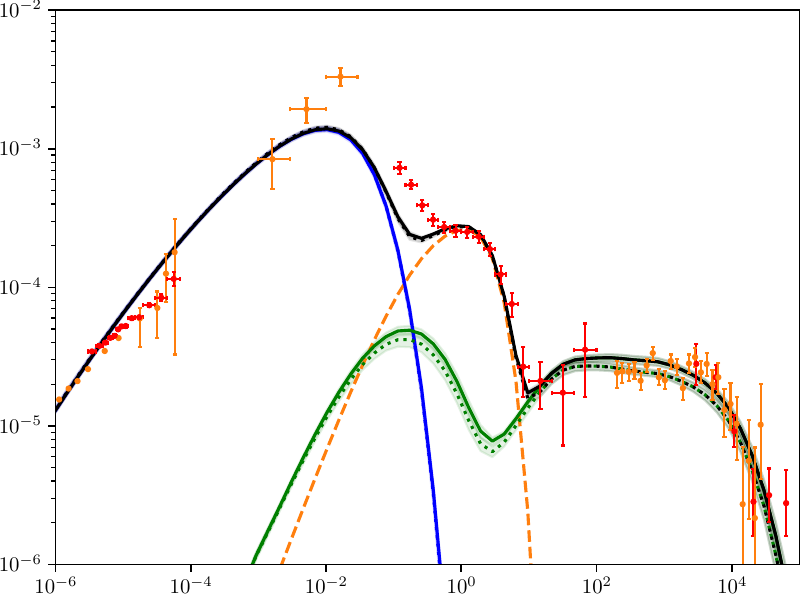}
        \end{picture}
}
\caption{
\label{FigSpectrumNonThermalModified}
Spectral energy distribution of the non-thermal emission predicted in our model for LS 5039 with the total emission power reduced by a factor of 2 for both \textsc{Infc} phase intervals covered in our simulation.
Here, we show synchrotron emission in blue and IC emission in green.
Results for the first full orbit are shown as solid lines and those for the second full orbit as dotted lines.
Additionally, the outer-gap model is shown in dashed orange.
The observational data are the same as in Fig.~\ref{FigSpectraINFCandSUPC}.
}
\end{figure}

In Fig.~\ref{FigSpectraINFCandSUPC}, we can see that  VHE emission is apparenty overestimated in the \textsc{Infc} phase interval by approximately a factor of 2.
Unfortunately, rerunning the model with a reduced injection factor is prohibitive due to the very high numerical cost; particularly  since the injection is regulated via three parameters in total (see Sect. \ref{SecInjection} and the related discussion in \citet{HuberEtAl2021AnA646_A91}).
This leads to a non-linear dependence of the normalisation of the injected spectra, especially for the Maxwellian.
Apart from that, it is not clear at this point whether the number of injected particles is too large in total or whether we would just have to shift either particles or energy from the power-law component to the Maxwellian one (regulated via $\zeta_n$ and $\zeta_e$ in Sec. \ref{SecInjection}).

\begin{figure}
\resizebox{\hsize}{!}{
        \setlength{\unitlength}{0.0081cm}
        \begin{picture}(1100,961)(-100,-50)
                \put(-100,50){\small \rotatebox{90}{$\displaystyle \frac{F_{> \text{1 TeV}}}{10^{-8} \text{m}^{-2} \text{s}^{-1}}$}}
                \put(-100,350){\small \rotatebox{90}{$\displaystyle \frac{F_{\text{0.2 - 3 GeV}}}{10^{-3} \text{m}^{-2} \text{s}^{-1}}$}}
                \put(-100,620){\small \rotatebox{90}{$\displaystyle \frac{F_{\text{3 - 10 keV}}}{10^{-14} \text{J\,m}^{-2} \text{s}^{-1}}$}}
                \put(450,-50){\small $\phi_\text{orbit}$}
                \includegraphics[width=1000\unitlength]{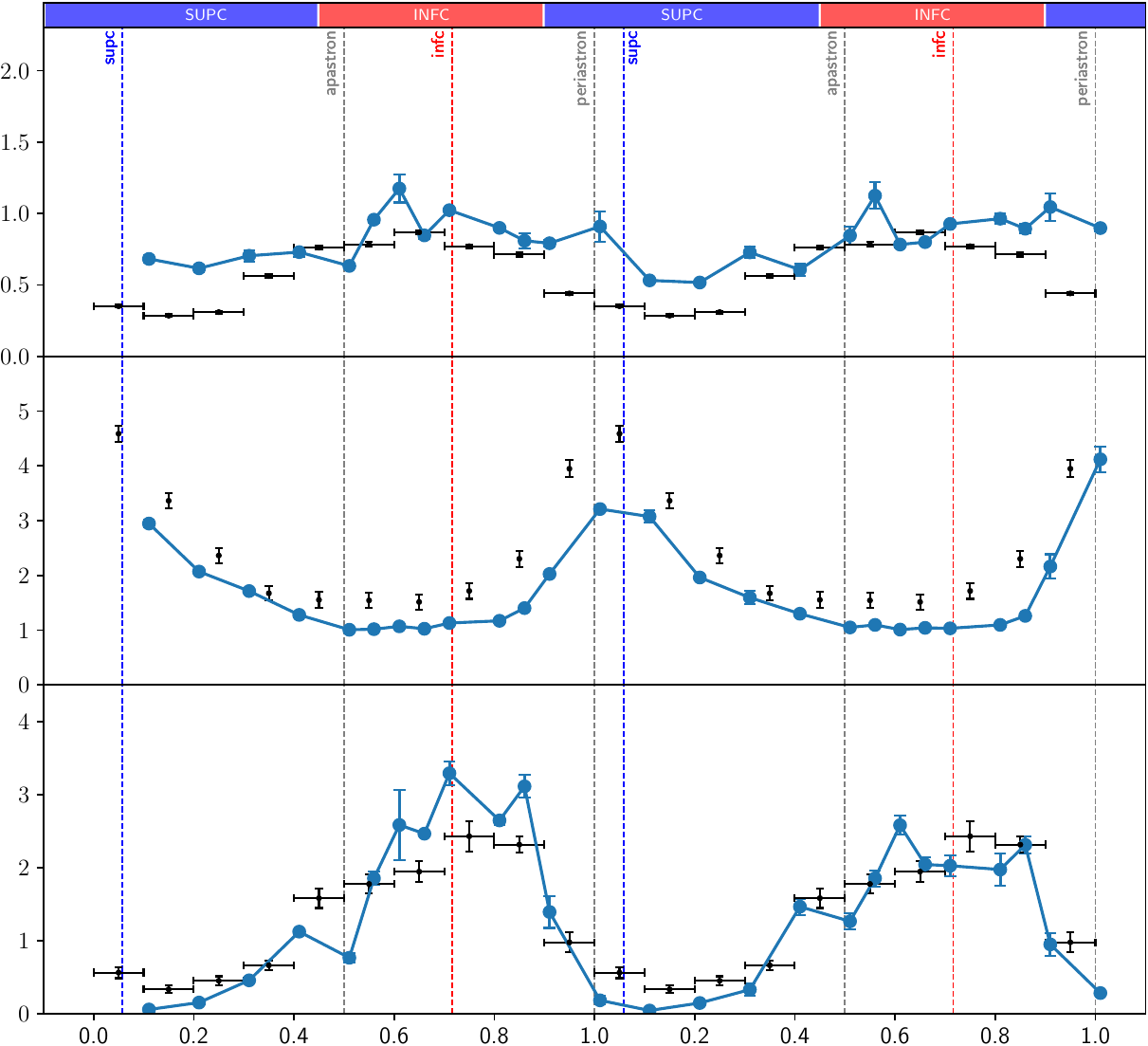}
        \end{picture}
}
\caption{\label{FigLightCurveModified}
Light curve of LS 5039 in three different energy bands with the total emission power predicted by the model reduced by a factor of 2.
Here, we show the model results in blue and corresponding data in black together with error bars where available.
Our simulation results are given for an inclination of $i=60^{\circ}$.
The individual energy bands are soft x-rays at an energy of 3-10\,keV (top row, with data from \citet{YonedaEtAl2020PhRvL125_1103}, HE gamma rays at an energy of 0.2-3\,GeV (middle row, with data from \citet{HadaschEtAl2012ApJ749_54}), and VHE gamma-rays with energies above 1\,TeV (lower row, with data from \citet{AharonianEtAl2006AnA460_743}).
}
\end{figure}
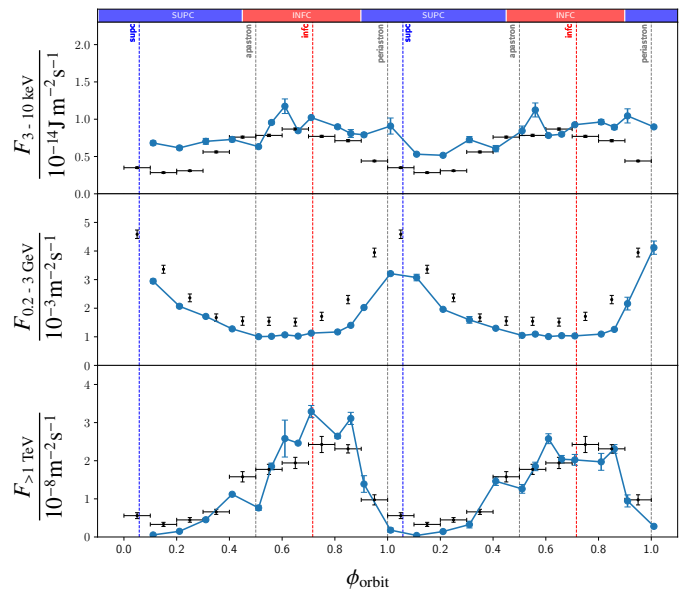

To test the impact of a modified injection, we simply reduced the non-thermal emission from the transported non-thermal leptons for the inclination $i=60^\circ$ by a factor of 2.
Here, we did not change the model for the magnetospheric emission, because this is not affected by the particle injection at the shocks.
We show the results of the corresponding SED for the \textsc{Infc} phase interval in Fig.~\ref{FigSpectrumNonThermalModified} and for the light curve in Fig.~ \ref{FigLightCurveModified}.
For the SED, we show the results for both simulated orbits together with error bands representing the standard deviation of the spectra used to compute the mean spectrum for each full \textsc{Infc} phase interval.
Apparently, the orbit-to-orbit variation in this case is consistent with this standard deviation.
With this modified emission model, we obtained an excellent fit for the SED during the \textsc{Infc} phase intervals above an energy of 1~GeV, especially with a significant improvement in the regime above 10~GeV.
The only issue in the TeV regime is the clear cut-off on our model above $\sim$10~TeV, whereas recent data indicate a continued power law to even higher energies \citep[see][]{AlfaroEtAl2025ApJ987L_42}.
This could possibly be remedied by an additional hadronic components (see the discussion below).

In addition, the modified light curves show a better correspondence to the observations.
In the VHE regime, they follow the data rather closely with the peak after apastron being strongly driven by relativistic beaming, which in the unmodified picture in Fig.~\ref{FigLightCurvesFullAmplitude} led to strong overshooting.
In the HE regime, while showing a vast improvement compared to our previous model in \citet{HuberEtAl2021AnA649_71}, our prediction is still below the observations.
As discussed above, this seems to be related to the low emission power of the Maxwellian, which is actually further suppressed by the additional flux reduction.

Finally, in the soft x-ray regime, we also find a strong improvement both compared to our previous model and compared to the unmodified light curve.
While we should certainly not overstress these findings given our simplified magnetic-field model, at least we are able to partially recover the general trend of the light curve in this regime.
This is probably related to the problem that in the previous model the Coriolis shock was not captured by the simulation any more in part of the \textsc{Infc} range.
Additionally, we see an impact of relativistic beaming for both orbits around orbital phase 0.6 (see also the discussion at the end of Sect. \ref{SecEmissionNonMod}).
At the same time, we observe some flux variability at these orbital phases (given by the standard deviation from the six data points distributed over 40 minutes, binned together in this case).
This is in accordance with our previous study of the wind dynamics in LS 5039 \citep[][]{KissmannEtAl2023AnA677_A5}, where we found strong variability of the distance of the Coriolis shock around the same time.

This discussion indicates that in a future model we have to aim at a new tuning of the injection parameters to improve the resulting fit.
Here we mainly indicated how to improve the fit by removing power (either in the form of a reduced contribution of density or thermal energy) from the non-thermal power law.
Since, in this post-processing analysis, we could just reduce the total emitted power in the emission from the system, we were not able to move the power from the non-thermal particles to the Maxwellian.

Actually such a combined effort could further improve the fit by improving the properties of the transported Maxwellian.
As already discussed above, Fig.~\ref{FigSpectraINFCandSUPC} shows that the average Lorentz factor of the Maxwellian in our simulations is probably too high.
From the possible injection parameters, modifying $\zeta_n^\text{PL}$ would only have a negligible impact on the Maxwellian, because $\zeta_n^\text{MW} = 1 - \zeta_n^\text{PL}$ with $\zeta_n^\text{PL}\ll 1$.
Correspondingly, we have to adapt the other parameters, which both have a non-linear impact on the injected Maxwellian.
This shows that  further simulations would be required to find a suitable setup of these parameters.

\section{Summary and discussion}
\label{SecConclusion}
In this study, we investigated the non-thermal emission predicted for our high-resolutions RHD plus particle-transport simulation of the LS 5039 system, where we assumed a wind-driven scenario to explain the acceleration of the non-thermal particles.
This study is an extension of our previous work \citet{HuberEtAl2021AnA649_71}, where we feature a higher spatial resolution and a significantly larger spatial domain.
As a consequence, we have been able to fully encompass all relevant structures for particle acceleration in our model.
The Coriolis shock, in particular, is contained within the numerical domain at all times.
As a result, we find a significantly higher total flux of non-thermal particles at the highest energy for phases after apastron.
Since we used the same injection parameters as in the previous model, we ended up overshooting the observed SED of the system by approximately a factor of 2 when assuming an inclination of $i=60^\circ$.
After reducing the emission correspondingly, we ended up with a better fit, with respect to both the SED and the light curve, than in the previous model. This result  hints at a preferred inclination of $i=60^\circ$.
However, this flux reduction had to be implemented into the post-processing procedure and, thus, it is not fully consistent.
Instead, it would be desirable for a future model to adapt the properties of injection at the shocks within the system consistently,  offering the possibility to further improve the quality of the match to observations.

The presented model is based upon the wind-driven scenario, namely, leptons from the highly relativistic pulsar wind are accelerated at shocks in the WCR.
The main acceleration sites are the bow shock, the Coriolis shock, and localised shock structures in the tail of the WCR.
Due to energy losses related to synchrotron and IC emission, the maximum energy of the non-thermal particles is limited, depending on the distance of the acceleration site from the stellar objects.
Especially around periastron, the acceleration sites are rather near to the stellar objects.
Consequently, the VHE gamma-ray emission is suppressed at these orbital phases.
Overall, we found a cut-off of the energetic-lepton spectrum in our simulation around a few tens of TeV.
This also translates to a similar cut-off in the IC photon spectrum \citep[see e.g.][]{LefaEtAl2012ApJ753_176}.
Here, we found a rather smooth cut-off in the photon spectrum because it is produced by a range of particle populations with cut-offs at different energies.
On the one hand, we have a large contribution from regions close to the stellar objects where the gamma-ray emission is high due to the high photon density, while also featuring a cut-off at lower lepton energies.
On the other hand, we have a large volume of outer regions with a cut-off of the lepton spectrum at higher energies, but with a smaller density of stellar photons.

As a result, we found the SED of non-thermal photons shows emission up to a few tens of TeV.
While this is consistent with the VHE emission detected by H.E.S.S., the recent measurements by HAWC indicate a continuous power-law spectrum for LS 5039 even beyond 100~TeV \citep[][]{AlfaroEtAl2025ApJ987L_42}.
Such high energies cannot be reached in our purely leptonic model since energy losses are too high at the main acceleration sites.
This also means that in the context of a wind-driven model, a power law exceeding 100~TeV has to be explained by an additional hadronic component.
For future modelling efforts, this means incorporating additional acceleration of protons at the shock of the stellar wind.

In conclusion, in our new RHD plus particle-transport model of LS 5039, we find an improved correspondence to observations of the emission from this system in the x-ray and gamma-ray regime.
Future numerical models should address the description of the magnetic field to improve the representation of synchrotron emission.
In addition, we found it is necessary to revise our setup regarding injection of the non-thermal power law and the transported Maxwellian.
A corresponding set-up should also consider additional updates of the injection model for the non-thermal particles. For instance, injection should depend on the compression ratio \citep[see][]{JonesEtAl1999ApJ512_105, VaidyaEtAl2018ApJ865_144V, WinnerEtAl2019MNRAS488_2235}, which might also improve the fit in the synchrotron regime.
Finally, it will be worthwhile to consider an additional hadronic component for the explanation of gamma-ray data beyond 100~TeV in future studies.

\begin{acknowledgements}
We thank N. Meindl for contributions to the gamma-ray analysis code and J. Takata for helpful discussions regarding the outer-gap model of LS 5039.
        Additionally, we thank S. Casanova for  providing us with the data by HAWC and H. Yoneda for providing us with the data by NuSTAR and Fermi.
        We thankfully acknowledge PRACE for granting us access to Joliot-Curie at GENCI@CEA, France. 
        We thankfully acknowledge the access to the research infrastructure of the Institute for Astro- and Particle Physics at the University of Innsbruck (Server Quanton AS-220tt-trt8n16-g11 x8).
  This research made use of Cronos \citep{KissmannEtAl2018ApJS236_53}; GNU Scientific Library (GSL) \citep{galassi2018scientific};  matplotlib, a Python library for publication quality graphics \citep{Hunter2007}; Scipy \citep{2020SciPy-NMeth}; and NumPy \citep{2020NumPy-Array}.
  We thank the referee D. Mukherjee for the helpful comments and suggestions that allowed us to improve our manuscript.
\end{acknowledgements}

\bibliography{referencesGamma}

\begin{appendix} 
\section{Analysis regions}
\label{AppendAnalysisRegions}
\begin{figure}
\resizebox{\hsize}{!}{
\setlength{\unitlength}{0.0081cm}
\begin{picture}(1040, 958)
\put(540,1){\small $x/$AU}
\put(5,470){\small \rotatebox{90}{$y/$AU}}
\put(40,50){\includegraphics[width=1000\unitlength] {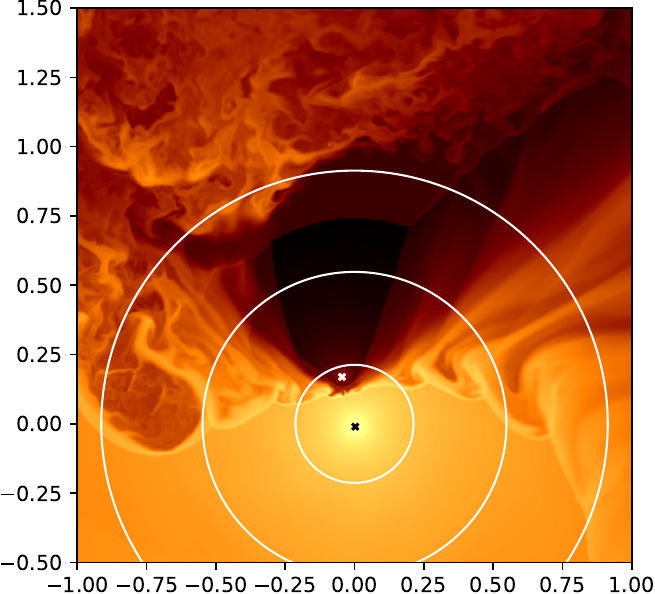}} 
\end{picture}
}
\caption{\label{FigAnalysisRegionsParticleSpectra}
Analysis regions for particle-spectra analysis overlaid over a plot of the mass density in the orbital plane in the vicinity of the pulsar cavity.
The positions of the early-type star and the pulsar are indicated via a black and a white cross, respectively.
The white circles indicate the different analysis regions for Fig.~\ref{FigParticleSpectrumContrained}.
These are given in order of increasing size: the inner region, the sheath region, and the Coriolis region.
}
\end{figure}

\begin{figure}
\resizebox{\hsize}{!}{
\setlength{\unitlength}{0.0081cm}
\begin{picture}(1040, 887.5)
\put(540,1){\small $x/$mas}
\put(5,440){\small \rotatebox{90}{$y/$mas}}
\put(40,50){\includegraphics[width=1000\unitlength]{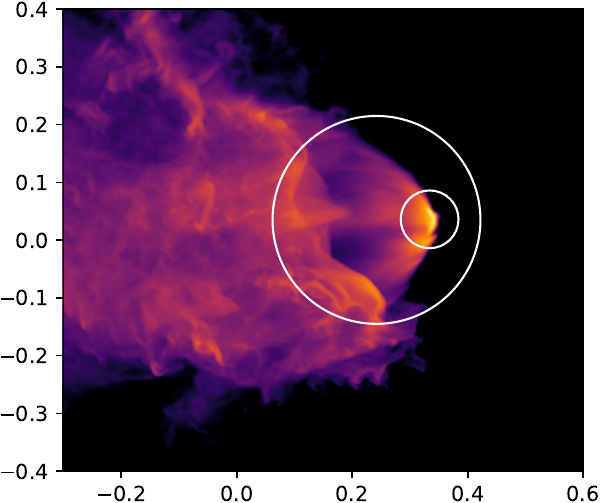}}
\end{picture}
}
\caption{\label{FigProjectionsMasks}
Analysis regions for emission-projection plots as shown in Fig.~\ref{FigProjections}.
Here, we show the corresponding circles overplotted over a projection for the synchrotron emission at 31.6~keV with an inclination of $i=60^\circ$.
The small white circle indicates the region, for which we summed up the emission around the bow shock.
The larger circle was used as an approximation for emission from the sheath around the pulsar, where the emission around the bow shock was subtracted.
}
\end{figure}

As an additional illustration, we show the different analysis regions in the figures in this section.
In Fig.~\ref{FigAnalysisRegionsParticleSpectra}, we show the analysis regions, used in producing the domain-integrated particle spectra for the different sub-regions in Fig.~\ref{FigParticleSpectrumContrained}.
For the sub-regions we used rings centred on the center of mass of the binary system as indicated in Fig.~\ref{FigAnalysisRegionsParticleSpectra}.
The sub-regions were chosen in a way, that the inner region contains the nose region of the bow shock, the sheath region contains the sheath of the WCR between bow shock and Coriolis shock, and the Coriolis region contains the vicinity of the Coriolis shock.
The rest of the domain is assigned to the outer region.
Additionally, Fig.~\ref{FigProjectionsMasks} shows the regions, over which we integrated the non-thermal flux for the different sub-regions discussed for Fig.~\ref{FigProjections}.

\section{Comparison to fluid quantities}
\label{AppendComp}
\begin{figure*}
  \setlength{\unitlength}{0.017143cm}
  \begin{picture}(1050,641)(0,0) 
  \put(30,0){\includegraphics[width=1000\unitlength]{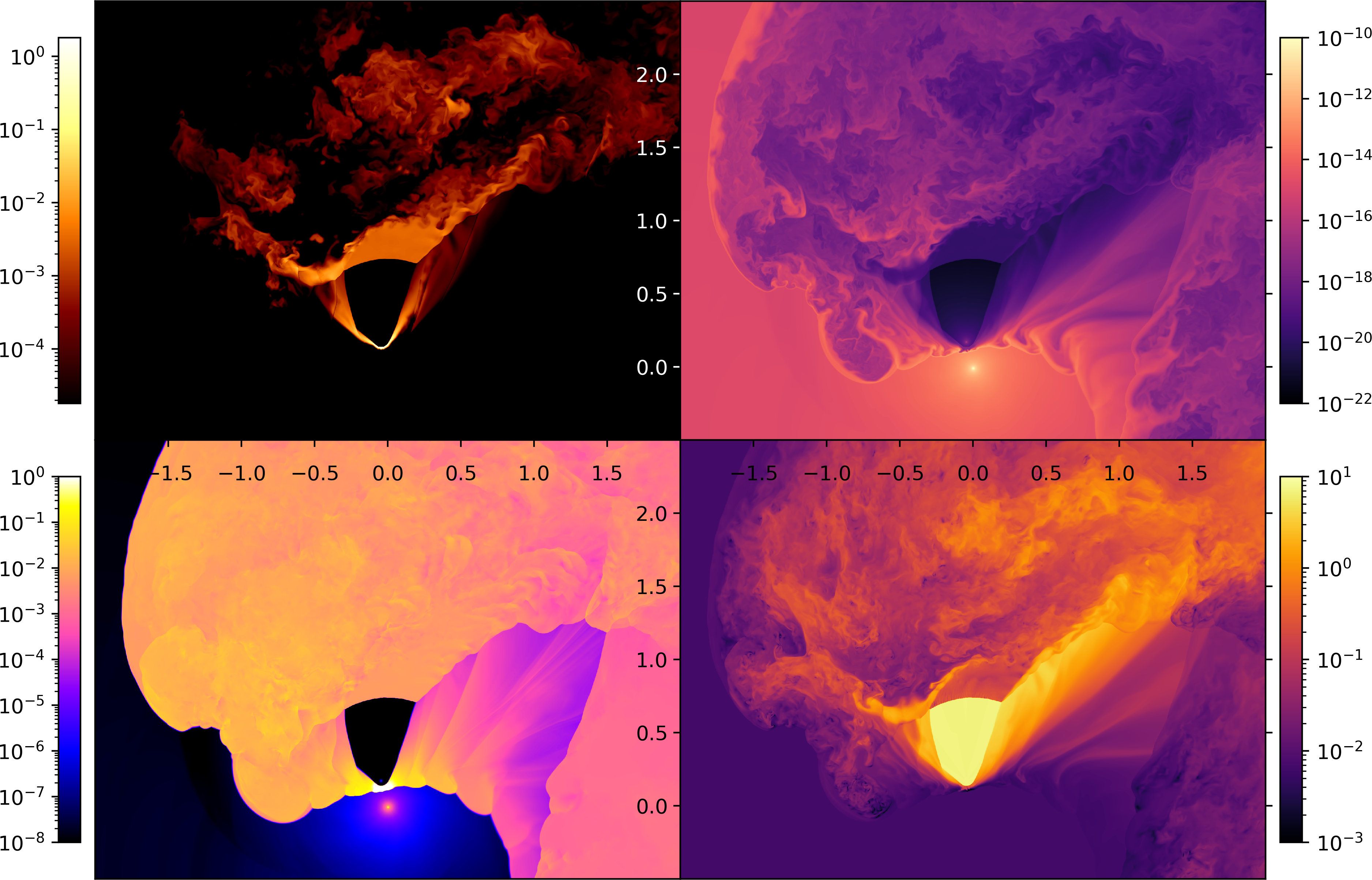}}
  \put(2,400){\rotatebox{90}{$n_l'(E = 485.5\ \text{GeV}) / \text{m}^3$}}
  \put(2,140){\rotatebox{90}{$p / \text{Pa}$}}
  \put(1035,450){\rotatebox{90}{$\rho / \text{kg}~\text{m}^3$}}
  \put(1035,150){\rotatebox{90}{$\left|u\right|$}}
  
  \put(470,140){\rotatebox{90}{$y/\text{AU}$}}
  \put(470,460){\textcolor{white}{\rotatebox{90}{$y/\text{AU}$}}}
  \put(290,270){$x/\text{AU}$}
  \put(720,270){$x/\text{AU}$}
  \end{picture}
\caption{\label{FigCompQuantities}
Number density of accelerated leptons shown in contrast with different fluid quantities in the orbital plane for an orbital phase of $\phi = 0.41$ during the second orbit.
We show the accelerated leptons on the upper left for an energy of 485.5\ GeV (see also Fig. \ref{FigElectronsPlane}). 
Additionally, we show the representative fluid quantities in the form of the mass density (upper right), the thermal pressure (lower left) and the absolute value of the spatial component of the fluid’s four velocity (lower right).
}  
\end{figure*}
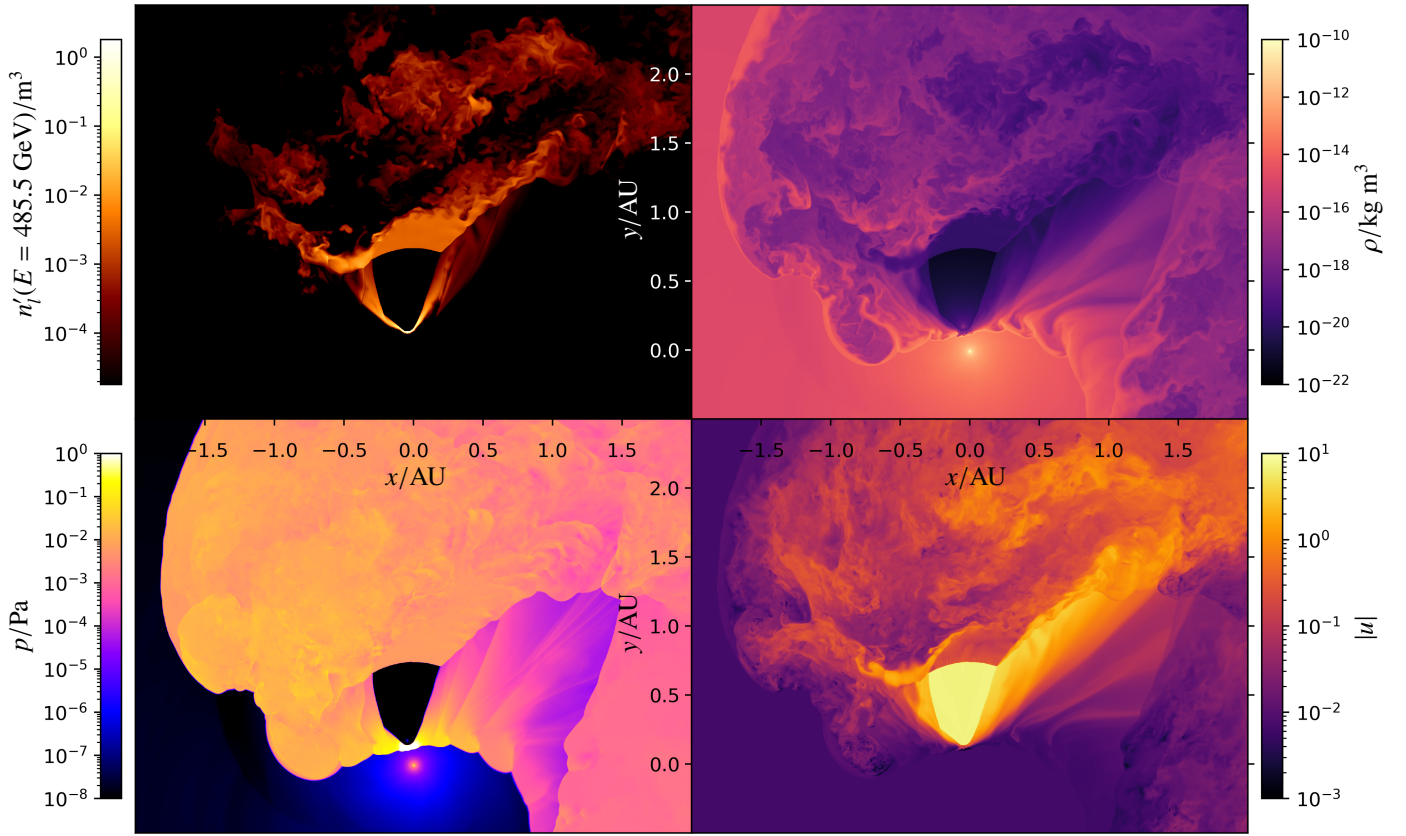
To relate the distribution of the number density of energetic leptons to properties of the interacting stellar and pulsar wind, we show the accelerated-lepton distribution together with the fluid quantities for the example of orbital phase $\phi = 0.41$ during the second orbit in Fig. \ref{FigCompQuantities}.
There, we show all results for the same time in the orbital plane.
From the comparison it becomes obvious that we only find energetic leptons in regions, where the fluid from the pulsar wind dominates.
This can be seen from the correlation between density of energetic leptons and four velocity or from the anti-correlation between the energetic lepton density and the mass density.
This behaviour follows because the energetic leptons are injected as a fraction of the pair plasma from the pulsar wind.

\section{Contribution of emission from different regions}
\label{AppendQuantContributions}
In this appendix, we supply some quantitative data for the contribution of different regions to the non-thermal emission of the simulated LS-5039 system.
For this, we analysed the projections at the different energies used to build the composite images in Fig.~\ref{FigProjections}.
The bow-shock region is defined as a small circular region around the bow shock with a radius of 0.05~mas.
The pulsar-sheath region is a circular region centred in the pulsar sheath with a radius of 0.035~mas, where we subtract the contribution from the bow-shock region.
Finally, the contribution of the tail region is computed by subtracting both other components from the total emission from the system.
Results for the synchrotron regime are given in Tables \ref{TabContrSynch30} and \ref{TabContrSynch60} for inclinations $i=30^\circ$ and $i=60^\circ$, respectively.
Similarly, Tables \ref{TabContrIC30} and \ref{TabContrIC60} show quantitative results for the emission in the IC regime for both inclinations.

\begin{table}
\caption{\label{TabContrSynch30}Contribution of emission regions in the synchrotron regime for an inclination of $i=30^\circ$.}
\centering
\begin{tabular}{c|ccc}
Region & $E=31.6$~keV & $E=1$~MeV & $E=31.6$~MeV \\
\hline
bow-shock & 55.3\% & 54.5\% & 68.6\%\\
pulsar-sheath & 15.6\% & 17.5\% & 13.3\%\\
tail region & 29.1\% & 28.0\% & 18.1\%\\
\hline
\end{tabular}
\end{table}

\begin{table}
\caption{\label{TabContrSynch60}Contribution of emission regions in the synchrotron regime for an inclination of $i=60^\circ$.}
\centering
\begin{tabular}{c|ccc}
Region & $E=31.6$~keV & $E=1$~MeV & $E=31.6$~MeV \\
\hline
bow-shock & 54.3\% & 53.7\% & 64.4\%\\
pulsar-sheath & 22.4\% & 25.8\% & 24.2\%\\
tail region & 23.3\% & 20.5\% & 11.4\%\\
\hline
\end{tabular}
\end{table}

\begin{table}
\caption{\label{TabContrIC30}Contribution of emission regions in the IC regime for an inclination of $i=30^\circ$.}
\centering
\begin{tabular}{c|ccc}
Region & $E=15$~GeV & $E=474$~GeV & $E=15$~TeV \\
\hline
bow-shock & 83.9\% & 68.2\% & 89.5\%\\
pulsar-sheath & 8.9\% & 15.2\% & 4.5\%\\
tail region & 7.2\% & 16.6\% & 6.0\%\\
\hline
\end{tabular}
\end{table}

\begin{table}
\caption{\label{TabContrIC60}Contribution of emission regions in the IC regime for an inclination of $i=60^\circ$.}
\centering
\begin{tabular}{c|ccc}
Region & $E=15$~GeV & $E=474$~GeV & $E=15$~TeV \\
\hline
bow shock & 78.4\% & 69.7\% & 86.6\%\\
pulsar sheath & 15.0\% & 19.1\% & 9.7\%\\
tail region & 6.6\% & 11.3\% & 3.7\%\\
\hline
\end{tabular}
\end{table}

\end{appendix}

\end{document}